\documentclass[12pt,oneside,letterpaper]{article}
\usepackage{etex} 
\usepackage{amssymb,amsfonts,amsmath,amstext}
\usepackage{natbib}
\usepackage{bibentry}    
\usepackage{setspace}
\usepackage{multirow}      
\usepackage{booktabs}      
\usepackage{bbm}           
\usepackage{enumitem}
\usepackage[utf8]{inputenc}

\usepackage{graphicx} 

\usepackage{tikz}
\usetikzlibrary{arrows,shapes,positioning,backgrounds,calc,trees}
\pgfdeclarelayer{background}
\pgfsetlayers{background,main}

\usepackage[amsmath,hyperref,framed,standard]{ntheorem} 
\theoremclass{Theorem}
\theoremstyle{plain}
\theoremheaderfont{\normalfont\bfseries}
\theorembodyfont{\itshape}

\renewtheorem{theorem}[Theorem]{Theorem}
\renewtheorem{definition}[Definition]{Definition}
\renewtheorem{lemma}[Lemma]{Lemma}
\renewtheorem{proposition}[Proposition]{Proposition}
\renewtheorem{remark}[Remark]{Remark}
\renewtheorem{corollary}[Corollary]{Corollary}

\theoremclass{Assumption}
\newtheorem{assumption}{Assumption}

\newcommand{\qed}{\hfill\rule{0.4em}{0.4em}}

\usepackage{rotating} 
\usepackage[margin=1.25in]{geometry}
\newcommand{\comments}[1]{}

\newcommand{\Pfn}{\mathcal{P}}               
\newcommand{\Gn}{\mathbf{S}}                 
\newcommand{\SSn}{\mathbf{S}_n}                 
\newcommand{\Sn}{\mathbf{S}}                 
\newcommand{\Si}{\mathbf{S}_{(i)}}           

\newcommand{\setI}{I}                       
\newcommand{\Ik}{I_k\backslash i}           

\newcommand{\Xn}{\mathbf{X}}        
\newcommand{\Smode}{S^{mode}}       

\newcommand{\neksn}{\textrm{NE}k\textrm{SN}}

\newcommand{\NS}{\mathbf{N}}        
\newcommand{\Ns}{\NS}        
\newcommand{\MK}{\mathbf{M}}        
\newcommand{\KK}{\mathbf{K}}        

\newcommand{\R}{\mathbb{R}}

\newcommand{\e}{\mathrm{e}}

\newcounter{countnoteitem}

\newcommand{\fignotetitle}[1]{\small \textit{#1}}
\newcommand{\fignotetext}[1]{\small {#1}}
\newcommand{\axislabel}[1]{\footnotesize \textit{#1}}

\definecolor{UniBlue}{RGB}{83,121,170}
\definecolor{lavander}{cmyk}{0,0.48,0,0}
\definecolor{violet}{cmyk}{0.79,0.88,0,0}
\definecolor{burntorange}{cmyk}{0,0.52,1,0}
\definecolor{rosybrown}{rgb}{0.74,0.56,0.56}

\def\nodeborder{orange}
\def\nodeborderTM{black}
\def\nodetext{black}
\def\nodefont{\bfseries\tiny}
\def\linkfont{\color{rosybrown} \scriptsize}
\def\node_index_color{\color{black}}
\def\inf_node_background_color{red!70}

\tikzstyle{node}=%
[draw,circle,inner sep=2.0pt,minimum width=5pt,
thick,\nodeborder,text=\nodetext,font=\nodefont]

\tikzstyle{inf node t}=%
[draw,circle,inner sep=2.0pt,minimum width=5pt,
thick,\nodeborder,text=\nodetext,font=\nodefont,
bottom color=\inf_node_background_color, top color=white]

\tikzstyle{inf node m}=%
[draw,star,star points=5,inner sep=0.8pt,minimum width=17pt,
thick,\nodeborderTM,text=\nodetext,font=\nodefont]

\tikzstyle{inf node tm}=%
[draw,star,star points=5,inner sep=0.8pt,minimum width=17pt,
thick,\nodeborderTM,text=\nodetext,font=\nodefont,
bottom color=\inf_node_background_color, top color=white]

\tikzstyle{inf node}=%
[draw,star,star points=5,inner sep=0.8pt,minimum width=17pt,
thick,\nodeborder,text=\nodetext,font=\nodefont,
bottom color=\inf_node_background_color, top color=white]

\tikzstyle{inf node a}=%
[draw,diamond,inner sep=0.8pt,minimum width=17pt,
thick,\nodeborder,text=\nodetext,font=\nodefont]

\tikzstyle{inf node b}=%
[draw,star,star points=5,inner sep=0.8pt,minimum width=17pt,
thick,\nodeborder,text=\nodetext,font=\nodefont]

\tikzstyle{legend_node}=%
[circle, draw, \nodeborder, rounded corners,thick,
bottom color=white, top color=white,
text=\nodetext,font=\tiny\bfseries,inner sep=0.0pt,text width=0.8cm,minimum width= 1.1cm,align=center]

\tikzstyle{legend_inf_node}=%
[star, star points=5, draw, \nodeborder, 
bottom color=\inf_node_background_color, top color= white,
text= \nodetext,font=\tiny\bfseries,inner sep=0.0pt,text width=0.6cm,minimum width= 0.9cm,align=center]

\tikzset{
    >=stealth',
    punkt/.style={
           rectangle,
           rounded corners,
           draw=black, very thick,
           text width=6.5em,
           minimum height=2em,
           text centered},
    pil/.style={
           ->,
           thick,
           shorten <=2pt,
           shorten >=2pt,}
}

\begin{document}

\title{Nash Equilibria on (Un)Stable Networks
\footnote{The latest version of the paper, an online appendix with robustness analysis, and the implementation code are available at \url{www.antonbadev.net/neks}.}
\footnote{Based on my Ph.D. dissertation (\citealp{badev2013discrete}) at the University of Pennsylvania under the guidance of Kenneth Wolpin, George Mailath and Petra Todd. I have benefited from discussions with Steven Durlauf, Hanming Fang, James Heckman, Matt Jackson, Ali Jadbabaie, Michael Kearns, Angelo Mele, Antonio Merlo, and \`{A}ureo de Paula, and from audiences at Bocconi, Cornell, 2012 SSSI (UChicago), 2012 QME (Duke), 2012 XCEDE, GWU, 2015 NSF-ITN (Harvard), Mannheim, Minnesota, 2012 NASM, NYU, Penn, Pitt/CMU, St. Louis (Econ and Olin), 2015 ESWC (Montreal), 2016 SITE (Stanford), 2014 SED (Toronto), 2014 EMES (Toulouse), Tilburg, Texas Tech, University of Hawaii, 2015 IAAE, 2016 AMES (Kyoto), and Yale (SOM) for useful comments.
I gratefully acknowledge financial support from the TRIO (PARC/Boettner/NICHD) Pilot Project Competition. This work used the XSEDE, which is supported by National Science Foundation grant number OCI-1053575. All errors are mine.%
}
\\
}
\author{Anton Badev%
\thanks{The views expressed herein are those of the author and not necessarily those of the Board of Governors of the Federal Reserve System.} \\
}
\date{ 
6/21/2020 
}
\maketitle

\vspace{-1.0cm}
\renewcommand{\abstractname}{}
\begin{abstract}
  \singlespacing
  \noindent \textbf{Abstract.}
  In response to a change, individuals may choose to follow the responses of their friends or, alternatively, to change their friends. 
To model these decisions, consider a game where players choose their behaviors and friendships. In equilibrium, players internalize the need for consensus in forming friendships and choose their optimal strategies on subsets of $k$ players - a form of bounded rationality. The $k$-player consensual dynamic delivers a probabilistic ranking of a game's equilibria, and, via a varying $k$, facilitates estimation of such games.

Applying the model to adolescents' smoking suggests that:
(a.) the response of the friendship network to changes in tobacco price amplifies the intended effect of price changes on smoking, 
(b.) racial desegregation of high-schools decreases the overall smoking prevalence, 
(c.) peer effect complementarities are substantially stronger between smokers compared to between non-smokers.

\end{abstract}

\newpage
\renewcommand{\thepage}{\arabic{page}}

\section{Introduction}
In response to a change in their environment, individuals may choose to follow the responses of their friends or, alternatively, choose to change their friends. In the context of evaluating public policies (e.g., an excise tax on tobacco consumption), these decisions motivate a shift from questions such as how the friendship network \emph{propagates} changes in individuals' behaviors, say, due to a policy change (e.g., an increase in tobacco price), to questions such as how the friendship network \emph{responds} to such changes in individuals' behaviors. This paper studies this shift in perspective from both a theoretical and public policy view.

In order to do so, consider an environment where individuals choose both their behaviors and friendships. While these choices are fundamentally different, their difference is not related to the presence of strategic incentives or instincts for selfish decisions. Rather, choosing a friend presumes a consent (\citealp{Jackson_Wolinsky_1996}) while choosing behaviors does not (\citealp{Nash1950}). The tension between the instinct for selfish choices and the consensual nature of humans' friendships can be prototyped as a game of link and node statuses where players' decision problem is augmented with a set of \emph{stability constraints}. These constraints reflect that a player internalizes the the need for consent in forming links, or in other words, a player chooses her links only among those who desire to be her friends.

Given a player's incentives, her observed friendship links and behaviors are likely to compare favorably against her alternatives, i.e., are likely to be robust against a set of feasible deviations. Reasoning about the complexity of individuals' decision problem\footnote{There are $2^{n-1}$ possible link deviations and only $n-1$ possible one-link deviations.}
motivates a family of equilibria indexed by the \emph{radius} of permissible deviations. For a fixed parameter $k$, a Nash equilibrium in a $k$-stable (NE$k$S) network emerges when no player has profitable deviation that is permissible by the stability constraints and that involves less than $k$ links. A feature of the proposed model is that NE$k$S networks are pairwise stable and, for $k=n$, NE$k$S networks are pairwise-Nash networks (see \citealp{Jackson2005survey} for an overview of these concepts).

A primitive of games of links and behaviors is payoff externalities %
 which can lead to multiplicity of NE$k$S networks. 
To reconcile this multiplicity, this paper introduces a $k$-player consensual dynamic ($k$CD)---a family of myopic dynamic processes where players sequentially adapt their behaviors and at most $k-1$ of their links, of course, subject to the stability constraints%
\footnote{Similar dynamics, although in a different context, are analysed by the evolutionary game theory and individual learning literatures, e.g., \cite{foster_young_90,kandori_mailath_rob_93,Blume_1993,JacksonWatts2001,JacksonWatts2002}. In a crude form, the motivation for these dynamics is present in \citet[Chapter VII]{Cournot_1838} and \cite[Section 9]{Nash1950}.}.
In the presence of random preference shocks, $k$CDs induce a unique, invariant to the choice of $k$, stationary distribution over the set of all possible outcomes. Intuitively, each NE$k$S network is a 
 local mode of this probability distribution.  
In addition, the larger $k$ is, the faster a $k$CD approaches the stationary distribution.

These properties of $k$CDs facilitate both estimation of and simulation from these games. The model's likelihood is given by the (unique) stationary distribution of the $k$CD family. This distribution pertains to the Exponential Random Graph Models (\citealp{FrankStrauss1986markov,WassermanPattison1996}), for which both direct estimation and simulating from the model with known parameters are computationally infeasible.%
\footnote{A likelihood evaluation involves summations with $2^{(n^2+n)/2}$ terms, e.g. for $n=10$, $2^{55}$ terms.} 
The double Metropolis-Hastings sampler offers a Bayesian estimation strategy that nevertheless relies on simulations from the stationary distribution via Markov chains (\citealp{murray2012mcmc,Liang2010double,Mele2017}). While, for different $k$s, $k$CDs have different convergence properties they have the same stationary distribution, which in turn suggests a transparent strategy for designing these Markov chains with \emph{varying} $k$.%
\footnote{Poor convergence properties are associated with local Markov chains, where each update is of size $o(n)$ (\citealp{BhamidiBrestlerSly_2011}). Importantly, varying $k$ on the support $\{2, \ldots, n-1\}$  is not anymore a local Markov chain. I thank an anonymous referee for pointing this out.}

The model is estimated with data on smoking behavior, friendship networks, and home environment (parental education background and parental smoking behavior) from the National Longitudinal Study of Adolescent Health.\footnote{Details about the Add Health data, including the sample construction, are in the appendix.} This is a longitudinal study of a nationally representative sample of adolescents in the United States, who were in grades $7$--$12$ during the $1994$--$95$ school year. 

The empirical models in \cite{Nakajima2007} and \cite{Mele2017} inspired the proposed framework. \cite{Nakajima2007} studies peer effects abstracting from friendships and \cite{Mele2017} obtains large network asymptotics of a model with link formation only. The approaches in these papers are fundamentally compatible so these models can be unified in a joint model, as in \cite{Boucher2016} and \cite{hsieha2016network}. Compared to existing empirical frameworks (including \citealp{CanenTrebbiJackson2016,BoucherHsiehLee2019,BattagliniPatacchiniRainone2019}), this paper explicitly models the strategic incentives guaranteeing network stability in the sense of \cite{Jackson_Wolinsky_1996}.

The empirical analysis of friendship networks and smoking behaviors lends support to a host of results which are related to the large body of empirical work on social interactions and teen risky behaviors. Typically, empirical studies on peer effects either lack data on friendship network or take the friendship network as given.\footnote{See, for example, \cite{LiuPatacchiniZenou2014}, who distinguish between local aggregate and local average peer effects, and the references therein.}
 Also the approaches range from models that directly relate an individual's choices to mean characteristics of his peer groups (e.g., see \cite{powell_tauras_ross_05} and \cite{ali_dwyer_09}) to models with elaborate equilibrium micro-foundations, such as those in \cite*{BrockDurlauf2001,BrockDurlauf2007,Krauth2005b,Calvo-ArmengolPatacchiniZenou2009}. In terms of estimates, this paper makes the first step in explaining how not accounting for the response of these social network could bias the estimates.\footnote{It is difficult, if not impossible, to account the empirical contributions of the large literature on peer effects and teen risky behaviors. For a small sample of papers obtaining estimates of peer effects see \cite{ChaloupkaHenry1997}, \cite{ali2009estimating} and the references in \cite[Surgeon General's Report]{CDCSurgeonGeneral2000}.}
Similarly, this paper pioneers a mechanism capable of explaining the role of the school composition, or more generally the determinants of the social fabric, on teen risky behaviors. The possibility of such a role was theorized by \cite{graham2014complementarity} and experimentally discovered by \cite{CarrellSacerdoteWest2013}.%

\subsection{Conclusions from the empirical analysis}

The model is estimated under various restrictions on the parameter specification and on the data availability. Two observations merit noting at the estimation stage. First, the peer effect complementarities are substantially stronger between smokers compared to between non-smokers. The model's parametrization permits differentiating these externalities and the conclusion is notable from the parameter estimates. Second, lack of network data, which forces the estimation to suppress the local peer effect externalities, substantially biases downwards the price coefficient.

The obtained sets of estimates are used to perform counterfactual experiments under various estimation scenarios. The purpose of these experiments is to quantify the response of the friendship network to policies targeting adolescent smoking. A by-product of this analysis is an assessment of the bias in the model's predictions due to lack of network data or due to various miss-specifications.

 The first experiment asks whether this response is relevant for policies working through changes in tobacco prices. To motivate this exercise, compare how individuals respond to a price increase in fixed versus endogenous network environments. There are two effects to consider. The \emph{direct} effect of changing tobacco prices is the first order response and, intuitively, will be larger whenever individuals are free to change their friendships. That is, more individuals are likely to immediately respond to changes in tobacco prices provided they are not confined to their (smoking) friends. The \emph{indirect} (ripple) effect of changing tobacco prices is the effect on smoking which is due, in part, to the fact that one's friends have stopped smoking. Contrary to before, a fixed network propels further the indirect effect. In a fixed network, an individual who changes her smoking status is bound to exert pressure to her (fixed) friends who are most likely smokers. It is, then, an empirical question how these two opposing effects balance out. Simulations with the full model and with a model where the friendship network is kept fixed suggest that the direct effect dominates. In other words, following an increase in tobacco prices the response of the friendship network amplifies the intended reduction in smoking prevalence.

The second experiment asks whether school racial composition has an effect on adolescent smoking. When students from different racial backgrounds study in the same school, they interact and are likely to become friends. Being from different racial backgrounds students have different intrinsic propensity to smoke and the question is what is the equilibrium behavior in these mixed-race friendships: do those who do not smoke start smoking or those who smoke stop smoking? Simulations from the model suggest that redistributing students from racially segregated schools into racially balanced schools decreases the overall smoking prevalence. 

The last experiment simulates a small scale policy intervention targeting only a part of school's population. The policy is efficient so that those exposed to the treatment stop smoking. At the same time it is not feasible (too costly) to treat the entire school. In this experiment, the question is when treated individuals return, will their friends follow their example, i.e. extending the effect of the proposed policy beyond the set of treated individuals and thus creating a domino effect, or will their pre-treatment friends un-friend them? In essence, this is a question about the magnitude of the spillover effects and this study suggests that aggregate spillovers are roughly double compared to the scale of the policy.

\subsection{Related literature}

This paper studies and estimates a game on endogenous network where players choose both their behaviors (e.g., smoking) and friendship links. The proposed model can be restricted to a game played on a fixed network. These games date back to the physics literature of the $70$s and in economics have been analyzed with both discrete and continuous choices (e.g., see \citealp{JacksonZenou2015} and \citealp{GamesPlayedonNetworks2016} for surveys). Most of the empirically tractable games have been developed either in continuous settings (e.g., \citealp{BallesterCalvo-ArmengolZenou2006}, \citealp{BramoulleKrantonDAmours2014}, \citealp{Calvo-ArmengolPatacchiniZenou2009}) or, when data on the friendship network is not available, restricting the model further to where peer effects are measured via group averages (e.g., see \citealp{BrockDurlauf2001,BrockDurlauf2007}, \citealp{Nakajima2007}, and the survey in \citealp{BlumeBrockDurlaufJayaraman2015}). 

Symmetrically, the proposed model can be restricted to a network formation game (e.g., see \cite{Jackson2008} for a systematic textbook presentation). A large and growing body of studies on the economics of these games followed \cite{Jackson_Wolinsky_1996} who, in a departure from the traditional non-cooperative game paradigm, introduced the notion of pairwise stability. In this paper, the stability constraints guarantee that any NE$k$S play is pairwise stable and for $k=n$ such play is pairwise-Nash (see \citealp{Myerson1991,CalvoArmengol2004,GoyalJoshi2006,BlochJackson2006,BlochJackson2007} and the survey in \citealp{Jackson2005survey}).
 
A handful of theoretical papers consider both network formation along with other choices potentially affected by the network (see \citealp{goyal_vega-redondo_05,CabralesCalvoArmengolZenou2011,KonigTessoneZenou2014nestedness,Baetz2014,
LageraasSeim2016strategic,Hiller2017,Jackson2016Paradox}). 
Importantly, the theoretical frameworks available are meant to provide focused insights into isolated features of networks and deliver sharp predictions, while abstracting from players' heterogeneity and so are not easily adapted for the purposes of estimation.

Econometric models of networks and actions are proposed in \cite{goldsmith2013social}, \cite{HsiehLee2016social} and \cite{JohnssonMoon2019} where the decisions to form friendships influence the decision to engage in a particular activity. The focus of their research, however, is not on policy analysis nor on accounting for the possible endogenous response of the friendship network to changing the decision environment. 
 In contrast, the framework proposed in \cite{Boucher2016} is microfounded as a particular equilibrium in a non-cooperative model of friendships and behaviors. 
Related work by \cite{hsieha2016network} proposes a two-stage estimation procedure, with an application to R and D, which relies on conditional independence of links delivered by abstracting from link externalities. \cite{CanenTrebbiJackson2016} propose an empirically tractable framework, building on \cite{CabralesCalvoArmengolZenou2011}, where politicians choose both socialization and legislation efforts, and study bill cosponsorship in the U.S. Congress.%
\footnote{There are recent contributions to the econometrics literature which focus on link formation, though these are not easily extendable to include action choice as well, e.g., see \cite{Sheng2014structural}, \cite{ArunJackson2016}, \cite{Leung2014random}, \cite{dePaulaRichardsTamer2014identification}, \cite{Graham2014econometric}, \cite{Menzel2015strategic} and the reviews in \cite{Arun2015econometrics,dePaula2016Econometrics,Bramoulle2016Oxford}.}
Differently to this literature, the proposed model is founded on the explicite strategic incentives that guarantee network stability in the sense of \cite{Jackson_Wolinsky_1996}.

Finally, adaptive dynamic and potential function representation, as a dimensionality reduction tool, are  widely used in (algorithmic) game theory, computer science and in economics of networks for processes on fixed networks, for processes of link formation and, more recently, for combined processes, e.g. \cite{foster_young_90}, \cite{Blume_1993}, \cite{JacksonWatts2001,JacksonWatts2002}, \cite{Nakajima2007}, \cite{BramoulleKrantonDAmours2014}, \cite{BourlesBramoullePerez-Richet2017}, \cite{Mele2017}, \cite{Boucher2016}, \cite{HsiehLee2016social}. %
In contrast to this literature, this paper highlights a slightly different role of the potential function, namely, as a tool to justify the gravitation of a family of adaptive dynamics around the equilibria of the static links and behaviors game. Further, the analysis of the $k$CD family justifies a model-based approach to simulate from and estimate these processes.



\section{A game on an endogenous network} \label{section:theory}
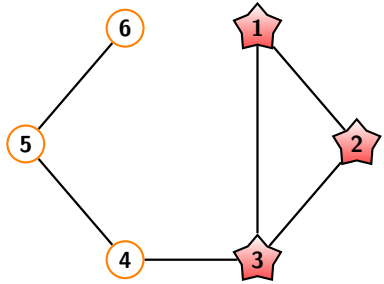
\begin{figure}[t]
\begin{center}
\caption{An example}
\vspace{0.5cm}
\def\node_index_color{\scriptsize\sffamily\bfseries\color{black}}
\def\linkfont{\color{white} \scriptsize}
\begin{tikzpicture}[->,scale=1.1,auto,every node/.style={scale=1.1},node distance=3cm]
\foreach \place/\name in {{(-2,0)/5},{(-0.8,1.4)/6},{(-0.8,-1.4)/4}}
\node[node] (\name) at \place   {\node_index_color\name};

\node[inf node tm] (1) at (0.8,1.4)  {\node_index_color 1};
\node[inf node tm] (2) at (2,  -0)   {\node_index_color 2};
\node[inf node tm] (3) at (0.8,-1.4) {\node_index_color 3};

\draw[-,line width=.3mm] (1) -- (2) ;
\draw[-,line width=.3mm] (1) -- (3) ;
\draw[-,line width=.3mm] (2) -- (3) ;
\draw[-,line width=.3mm] (3) -- (4) ;
\draw[-,line width=.3mm] (4) -- (5) ;
\draw[-,line width=.3mm] (5) -- (6) ;
\end{tikzpicture}
\label{diag1}
\end{center}
\fignotetitle{Note:} \fignotetext{In the graph, each player in $I=\{1,2,3,4,5,6\}$ is depicted as a vertex and a friendship is depicted as an edge, e.g. $1$ and $2$ are friends. The star-shaped shaded nodes denote players who smoke tobacco, e.g. $1$, $2$, and $3$ are smokers.}
\end{figure}

Imagine a world populated by individuals who chose both their friends and their behaviors, e.g. to smoke or not. Figure \ref{diag1} provides an example with $6$ individuals. In the figure, individuals are depicted as nodes on a graph and the star-shaped shaded nodes are those who smoke. Next, friendships are depicted as links between pairs of nodes. These links are undirected because (being in) a friendship is a symmetric binary relation. 

Before introducing the details of the game, it is worth enumerating the distinct features of this decision environment that are also reflected in the model. First, player $i$'s choices of behavior and friendships are different in that friendships (unlike behaviors) require consent to form and maintain. Second, there are likely to be externalities not only between individual's behavior and the behaviors her friends but also between individuals' friendship decisions. These various types of externalities are explicitly defined in the proposed payoff specification. Finally, this is a complex decision environment in that even with 10 individuals, each player considers roughly $1000$ alternative strategies while with $100$ individuals each player considers roughly $10^{30}$ alternative strategies. This complexity relates to the proposed (family of) equilibria and adaptive dynamic.

The model is developed in two stages. First, agents' strategic behavior is analyzed in static settings. Then, section 3 develops a family of myopic dynamic processes used to approximate the predictions of the static model in a inferentially convenient way.

\subsection{Players and preferences}

Each $i$, in a finite population $I=\left\{1,2,...,n\right\}$, chooses $a_{i}\in \{ 0,1\}$ and a set of links $g_{ij}=g_{ji}\in \{ 0,1\}$ for $j\neq i$. In the settings of adolescents' smoking and friendship decisions, $I$ is the set of all students in a given high school, $a_{i}=1$ if student $i$ smokes, and $g_{ij}=1$ if $i$ and $j$ are friends. These are the settings in figure \ref{diag1} above, with $3$ smokers, e.g. $a_1=a_2=a_3=1$, and $6$ friendships, e.g. $g_{12}=1$, $g_{23}=1$, etc. 
A final piece of the description of the population is individuals' exogenous characteristics $X_i$, e.g. age, gender, etc. 

Individual $i$ chooses her behavior and friendship statuses $S_{(i)} = (a_i,\{g_{ij}\}_{j\neq i})$ from her choice set $\Si=\{0,1\}^n$ to maximize her payoff $u_i$. Let $S=(S_{(1)},\ldots,S_{(n)})\in \prod_i \Si = \Sn$ and $X=(X_1,...,X_n)\in \Xn$. Formally $i$'s payoff function, $u_i: \Sn \times \Xn\longrightarrow \R$, orders the outcomes in $\Sn$ given $X$:%
\begin{eqnarray} \label{payoff}
u_i (S,X) & = & a_i v_i 
+ \underbrace{a_i \phi \sum_{j\neq i} a_j}_\text{aggr. externalities} 
 \\ \label{payoff2}
& & 
+ \underbrace{ \phi_{S} \sum_j g_{ij} a_i a_j + \phi_{N} \sum_j g_{ij}(1-a_i)(1-a_j)}_\text{local externalities} \\ \label{payoff3}
& & + \sum_j g_{ij} w_{ij}
+ \underbrace{q_{ijk} \sum_{\substack{j,k\\j<k}} g_{ij}g_{jk}g_{ki}}_\text{clustering}
- \underbrace{ \psi \left( \frac{1}{2} (d_i^2 + d_i) + \sum_{\substack{j\neq i}}  g_{ij} d_{j} \right)}
_\text{(convex) $\mathop{cost}_{i}(d_i,\{d_j\}_{j\in d_i})$}
\end{eqnarray}
where $d_i=\sum_j g_{ij}$ is the degree (total number of links) of $i$.
Here $v_i = v(X_i)$, $w_{ij}=w(X_i,X_j)$ and $q_{ijk}=q(X_i,X_j,X_k)$ are functions of agents' (exogenous) characteristics. To avoid clutter in the summation ranges, assume that $g_{ii}$ is defined and equal to zero for all $i$ so that, for example, $d_i=\sum_{j\neq i} g_{ij}=\sum_j g_{ij}$.

Note how $u_i$ depends both on individual's exogenous characteristics $X_i$ (e.g., terms $v_i$ and $w_{ij}$) and on her endogenous characteristics, e.g. number of friends, smoking statuses of her friends, and etc. More specifically, the terms in payoff (\ref{payoff}-\ref{payoff3}) can be sorted into three groups: terms that relate to the incremental payoff of changing $a_i$, terms that relate to the incremental payoff of changing $g_{ij}$ and terms that relate to both.

The first three terms in (\ref{payoff}-\ref{payoff3}) relate to the incremental payoff of changing $i$'s behavior $a_i$ conditional on the friendship network,
\begin{equation*}
\Delta_{a_i}u_i(S,X) = v_i + \phi \sum_{j\neq i} a_j 
+ \phi_S \sum_{j\neq i} g_{ij} a_j - \phi_N \sum_{j\neq i} g_{ij} (1-a_j).
\end{equation*}
The first term $v_i$ is the (exogenous) intrinsic utility of choice $a_i=1$ which is allowed to vary with $i$'s attributes $X_i$. The second term $\phi \sum_{j\neq i} a_j$ captures the aggregate externalities. That is, $i$ may be influenced from the behaviors of the surrounding population $\sum_{j\neq i} a_j$, provided $\phi\neq 0$. The last two terms in $\Delta_{a_i}u_i(S,X)$ are the differential of the local externalities $\phi_S \sum_j g_{ij} a_i a_j + \phi_N \sum_j g_{ij} (1-a_i)(1-a_j)$ in \eqref{payoff2}. Note that $a_i a_j$ equals $1$ if and only if $a_i=a_j=1$ so that, conditional on the friendship network, this term captures pressures on $i$ to follow (or to break away if $\phi_S<0$) her friends' decision to chose $1$ (to smoke). Analogously, $(1-a_i)(1-a_j)$ equals $1$ if and only if $a_i=a_j=0$, and this term captures pressures on $i$ to conform to the behaviors of her choosing $0$ (non-smoking) friends. Because $\phi_S$ need not equal $\phi_N$, the opposing conformity pressures from friends who choose $1$ and from friends who choose $0$ need not be equal in magnitude. Finally, as will become evident shortly, the local externalities terms are related to the incremental payoff of changing $g_{ij}$ where, conditional on individuals' actions, these terms capture a tendency to befriend others playing the same action. To sum up, an agent's utility increases by $\phi_S$ with every friend who plays the same action if that action is $1$, and by $\phi_N$ with every friend who plays the same action if that action is $0$.

The last four terms in (\ref{payoff}-\ref{payoff3}) relate to the incremental payoff to $i$ of changing $g_{ij}$ conditional on players' actions: 
\begin{eqnarray*}
\Delta_{g_{ij}}u_i(S,X)
&= & w_{ij} + q_{ijk}\sum_k g_{ik}g_{jk} - \psi (d_i+d_j) 
 + \phi_S a_i a_j + \phi_N(1-a_i)(1- a_j).
\end{eqnarray*}
The first term $w_{ij}$ captures the (exogenous) utility of a friendship which may depend on $i$'s and $j$'s degree of similarity, i.e., same sex, gender, race, etc. The next term is the differential of $q_{ijk}\sum_{j<k} g_{ik}g_{jk}g_{ki}$ in \eqref{payoff3} which captures link externalities. Mechanically, $i$ may have preferences for whether or not her friends are friends themselves. In particular, $i$ may prefer sharing her friends ($q>0$) or, on the contrary, prefer friendship exclusivity ($q<0$).%
\footnote{A compelling interpretation of this term is consistent with the presence of meeting frictions. In particular, meeting and befriending friends of friends can explain the tendency of individuals to form triangles of friendships (e.g., see \citealp{JacksonRogers2007}). This paper studies relatively small friendship networks so frictions are less likely to play a pronounced role.} 
 The third term is the differential of the convex cost term in \eqref{payoff3} which reflects the costs of establishing a friendship between $i$ and $j$. Properties of the cost term to note are: (i.) the more friends $i$ has, the more costly it is for $i$ to establish an additional friendship and (ii.) the costs are shared so for $i$ it is more costly to maintain friendships with more popular (high $d_j$) as opposed to less popular (low $d_j$) individuals. The last two terms relate to the previously discussed local externalities terms $\phi_S \sum_j g_{ij} a_i a_j + \phi_S \sum_j g_{ij} (1-a_i)(1-a_j)$ in \eqref{payoff2}.


\subsection{Equilibrium play}

Given a player's preferences, her observed links and action are likely to compare favorably against her alternatives. However, the number of available alternatives renders players' decision problem complex\footnote{There are $2^{n-1}$ possible link deviations and only $n-1$ possible one-link-at-a-time link deviations.} and motivates a family of equilibria where players consider only strategies that are close by, or in other words, where players consider deviations that are small. Here, the notion of closeness naturally translates to strategies that involve changing only few links. A final point concerning the equilibrium play is that players are aware that links are formed with consent.

\begin{definition} \label{def:eqlb}
A profile of actions and a network $S^*=(\{a_i^*\}_{i\in \setI},\{g_{ij}^*\}_{i\in \setI, j\in \setI\backslash i})$ is a \textbf{\emph{Nash equilibrium in a $k$(-player) stable (NE$k$S) network}}, provided $S^*_{(i)} = (a^*_i,\{g^*_{ij}\}_{j\neq i})$ is a solution of $i's$ decision problem on $I_k\subseteq I$:
\begin{eqnarray}\label{iproblem}
& \mathop{\mathrm{max}}_{a_i,\{g_{ij}\}_{j\in \Ik}} & u_i(a_i,\{g_{ij}\}_{j\in I\backslash i}; S_{-i}^*)
\\ \label{stability_constraint}
& \mathop{\mathrm{s.t.}} \quad &
\begin{array}{ccc}
g_{ij}=1 & \text{\normalfont only if} & \Delta_{g_{ij}}u_j(a_i,\{g_{ij}\}_{j\in I\backslash i}; S_{-i}^*) \geq 0  \quad \forall j\in \Ik
\end{array}
\end{eqnarray}
where $1<k\leq n$, $I_k=\{i\} \cup \{i_1,\ldots,i_{k-1}\}$ and $i\notin \{i_1,\ldots,i_{k-1}\}$, for all $i$ and $I_k$.

\end{definition}

To state the above definition in words, in a NE$k$S network no player has permissible, by the stability constraints \eqref{stability_constraint}, and profitable deviation involving changing the statuses of less than $k$ links. A notable feature of the NE$k$S networks is that not only links are formed with consent but also players internalize the need for consent through subjecting their play to the stability constraints. The stability constraints owe their name to their relation to the notion of stability introduced in \cite{Jackson_Wolinsky_1996} (see proposition \ref{prop:neksn-prop} below).

\begin{assumption} \label{asm:potential}
Assume that $w()$ and $q$ are symmetric in their arguments/indices.
\end{assumption}

\begin{proposition}\label{prop:neksn-exist}
With the utilities in \eqref{payoff}
\begin{enumerate}
\item For any $S$, $k$, $i$ and $I_k$, the problem in (\ref{iproblem}-\ref{stability_constraint}) is well defined and has a solution.
\item For any $k$, a NE$k$S network exists.
\end{enumerate}
\end{proposition}

The existence of a solution to the individual's decision problem in (\ref{iproblem}-\ref{stability_constraint}) and an equilibrium follows from the existence of potential function for this game (\citealp{monderer_shapley_96}). The proof is in appendix A (p. \pageref{proof:prop-neksn-exist}).

\begin{proposition}\label{prop:neksn-prop}
With the utilities in \eqref{payoff}
\begin{enumerate}
\item For $k=2$, NE$k$S networks are pairwise stable;
\item For $k=n$, NE$k$S networks are pairwise-Nash networks;
\item For $k'<k$, any NE$k$S network is also a NE$k'$S network.
\end{enumerate}
\end{proposition}

Part 1 can be strengthen for any preferences: for $k=2$, any NE$k$S play is pairwise stable (\citealp{Jackson_Wolinsky_1996}). For $k=n$, NE$k$S networks are pairwise-Nash networks (\citealp{CalvoArmengol2004,GoyalJoshi2006,BlochJackson2006,BlochJackson2007}). Finally, the NE$k$S family is ordered by set inclusion so that the existence of a pairwise stable network is a necessary condition for the existence of a NE$k$S network. The proof is in appendix A (p. \pageref{proof:prop-neksn-prop}).

\subsection{An example}

\begin{figure}[t!]
\begin{center}
\caption{Examples of NE$k$S networks for $k=2$ and $k=3$}
\label{fig:example-neks}
\vspace{0.5cm}
\includegraphics[scale=1.1,angle=0]{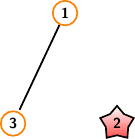}
\hspace{1.0cm}
\includegraphics[scale=1.1,angle=0]{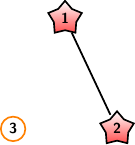}
\label{example_neks}
\end{center}
\fignotetitle{Note:} \fignotetext{Let $I=\{1,2,3\}$ and $u_i=v_i + \phi \sum_j g_{ij} (a_i a_j + (1-a_i)(1-a_j)) - \psi \mathop{cost}_{i}(d_i,\{d_j\}_{j\in d_i})$. It is straightforward to construct examples where for $k=3$ there is a unique NE$k$S network while for $k=2$ both of the depicted networks are NE$k$S networks.}
\end{figure}

To see how the choice of $k$ may affect the equlibrium networks, consider a simplified version of payoffs (\ref{payoff}-\ref{payoff3}) where all externalities other than the local peer effects and costs are absent. Let $I=\{1,2,3\}$, $\phi=0$,  $\phi_S=\phi_N=\phi_0$, and $q=0$. Also let $w_{ij}=0$ for all $i$ and $j$ so that 
\begin{equation}\label{example:neks}
u_i=v_i + \phi_0 \sum_j g_{ij} (a_i a_j + (1-a_i)(1-a_j)) - \psi \mathop{cost}_{i}(d_i,\{d_j\}_{j\in d_i}).
\end{equation}
Further, let $v_{2}=\bar{v}$ and $v_{3}=-\bar{v}$ for $\bar{v}$ large so that it is always a dominant strategy for players $2$ and $3$ to choose $a_2=1$ (smoke) and $a_3=0$ (not smoke) respectively. Finally, if it is costly to acquire friends ($\psi>0$) then players will never choose a friend playing different action because $w_{ij}=0$, and if the benefits of having a friend that plays the same action outweigh these costs ($\phi_0 > 2\psi$) then player $1$ would want to have exactly one friend (either player $2$ or player $3$) with the same smoking status. These candidates for equilibrium are depicted in figure \ref{fig:example-neks}.

For $k=3$ there is (generically) a unique NE$k$S network. If $v_1 < 0$ then player $1$ chooses not smoke and befriends player $2$ (figure \ref{fig:example-neks} left) else ($v_1 > 0$) player $1$ chooses to smoke and befriends player $3$ (figure \ref{fig:example-neks} right). In contrast, for $k=2$ if $v_1 \in \left(-\phi_0+2\psi,\phi_0-2\psi\right)$ both networks in figure \ref{fig:example-neks} are NE$k$S networks. Note that the larger the complementarities are ($\phi_0$), the larger the region for $v_1$ is where there are multiple NE$k$S networks.

\section{Consensual dynamic. An estimable framework}

The NE$k$S play offers an intuitive prescription for the \emph{outcomes} of the forces driving behaviors and friendships, without specifying the decision \emph{process} leading to these outcomes. This abstraction is challenged by strong informational assumptions where players are presumed to correctly anticipate other players' choices and by the presence of multiple NE$kS$ networks none of which can be ruled out a priori. Turning to a framework based on adaptive dynamics and random utility delivers a way to embed this multiplicity into an inferentially convenient framework.


Formulation (\ref{iproblem}-\ref{stability_constraint}) of individuals' decision problem  provides a basis for an adaptive process where behaviors and friendships evolve towards (or around) a NE$k$S network. The general idea that equilibrium might arise from simple (myopic) adaptive dynamic as opposed to from a complex reasoning process is very intuitive. The particular emphasis, compared to the interpretations in \cite{kandori_mailath_rob_93}, \cite{Blume_1993} and  \cite{JacksonWatts2001,JacksonWatts2002}, is on obtaining an estimable structure via a flexible adaptive process (parametrized via $k$).
\footnote{The literature on stochastic stability has studied stochastic dynamic where shocks vanish over time as a tool for equilibrium selection. In addition, a typical approach has been to analyze adaptive dynamics where either agents take turns to update their strategies (i.e., in our settings, all links) or a pair of players update the status of their link.
}
This flexibility presents advantages in simulating and estimating these games.

\subsection{$k$-player consensual dynamic ($k$CD)}

Every period $t=1,2,\ldots$ a randomly chosen individual, say $i$, considers $k-1$ of her friendships, say with $\{i_1,\ldots,i_{k-1}\}$, and her behavior $a_i$. In particular, $i$ myopically solves her decision problem (\ref{iproblem}--\ref{stability_constraint}) on $I_k=\{i\}\cup \{i_1,\ldots,i_{k-1}\}$. A stochastic meeting process $\mu_t$ outputs $i$ and $I_k$:
\begin{equation} \label{meeting}
\Pr \left( \mu _{t}=\left( i,I_k\right) |S_{t-1},X\right)
= \mu_{i,I_k} \left(S_{t-1},X\right)
\end{equation}
In the simplest case, when any meeting is equally probable, $\mu_{i,I_k} \left(S_{t-1},X\right)=\frac{1}{n} \frac{1}{{n-1 \choose k-1}}$ for all $i$, $I_k$, $S_{t-1}$, and $X$. However, we only need that any meeting is possible.

\begin{assumption} \label{asm:meet1}
$\mu_{i,I_k} \left(S_{t-1},X\right) > 0 $
for all $i \in I$, $I_k$, $S \in \Gn$ and $X \in \Xn$.
\end{assumption}

The sequence of meetings together with players' optimal decisions induce a sequence of network states $(S_t)$, which is indexed by time subscript $t$ and which will be referred to as \emph{$k$(-player) consensual dynamic} ($k$CD).

\begin{proposition} \label{prop:dynamics}
Fix $k\in[2, n]$. With assumptions \ref{asm:potential} and \ref{asm:meet1}, for a $k$CD $S_t$:
\begin{enumerate}
  \item Any NE$k$S network is absorbing, i.e. $S_{t'}=S_t$ if $S_t$ is a NE$k$S network, $t'>t$;
  \item Independently of the initial state
 $\Pr \left( \lim_{t\rightarrow \infty} S_t \in \neksn \right) = 1.$
\end{enumerate}
\end{proposition}

Indeed, for any $k$, the NE$k$S networks are exactly the rest points of simple decision processes, the $k$CDs. The proof is in Appendix A (p. \pageref{proof:prop-dynamics}).



\subsection{$k$CDs with random utility}

Consider the following modification of a $k$CD. In each period after the meeting is realized, the decision problem of the player who is drawn to make a choice is cast as a random utility choice.%
\footnote{This is also known as the random utility model. See \cite{thurstone_27}, \cite{marschak_60}, \cite{mcfadden_74} and, for textbook treatement, \cite[Chapter 2]{Train2003}.} That is, player's payoffs for each alternative are augmented with a random component, ultimately making her choice stochastic. Because players' choices are stochastic, such a $k$CD with random utility delivers a distribution over possible NE$k$S networks as opposed to a single network (see proposition \ref{proof:prop-dynamics}). Moreover, this distribution has some convenient properties when treated as (the) likelihood.

\begin{assumption} \label{asm:error_utility}
Suppose that the utilities in (\ref{payoff}) contain an additive random preference shock $u_i(S,X)+\epsilon_{S}$ where $\epsilon_{S} \sim i.i.d.$ across time and network states. Moreover, suppose that $\epsilon_{S}$ has c.d.f. and unbounded support on $\mathbb{R}$.
\end{assumption}

\begin{assumption} \label{asm:gumbel}
Suppose that the preference shock $\epsilon$ is distributed $Gumbel (\mu_\epsilon,\beta_\epsilon)$.
\end{assumption}

\begin{assumption} \label{asm:meet2}
Suppose that the meeting probability in \eqref{meeting}, $\mu_{i,I_k}(S,X)$ does not depend on $a_i$ and $g_{ij}$ for all $j\in I_k$. (Alternatively, which is slightly weaker, suppose that $\mu_{i,I_k}(S,X) = \mu_{i,I_k}(S',X)$ for all $S,S'\in \Gn$.)
\end{assumption}

The meeting process $\{\mu_t\}_{t=1}^{\infty}$ and the sequence of optimal choices, in terms of behaviors and friendship links, induce a Markov chain on $\Gn$ referred to as a $k$CD with random utility. The family of $k$CDs with random utility obey some desirable properties. (The proof is on p. \pageref{proof:thm-stationary}.)

\begin{theorem} \label{thm:stationary}
\textsc{[Stationary distribution]}
Fix $k\in[2,n]$. The $k$CD with random utility has the following properties:
\begin{enumerate}
 \item With assumptions \ref{asm:meet1} and \ref{asm:error_utility}, there is a unique stationary distribution $\pi_k \in \Delta(\Gn)$ for which $\lim_{t \rightarrow \infty} \Pr (S_t = S) = \pi_k(S)$. In addition, for any function $f:\Gn \rightarrow \mathbb{R}$, 
$\frac{1}{T} \sum_{t=0}^T f(S_t) \longrightarrow \int f\left( S\right) d \pi_k.$

\item With assumptions \ref{asm:potential}-\ref{asm:meet2},
\begin{equation}\label{logl}
 \pi(S,X) \propto \exp \left( \frac{\Pfn(S,X)}{\underline{\beta}} \right).
\end{equation}
In particular, $\pi(S,X)$ does not depend on $k$.
\end{enumerate}
\end{theorem}

The first part is not surprising in that it asserts that a $k$CD with random utility is well behaved so that standard convergence results apply. The uniqueness of $\pi_k$ precludes dependence between snapshots from this process and its initial state, and the ergodicity allows to simulate from $\pi_k$ via drawing a long trajectory of the $k$CD. 

The second part of the theorem has implications for implementing the model. Note how in (\ref{logl}) the stationary distribution $\pi$ does not depend on $k$ and, thus, delivers a tool to unify the equilibria in the NE$k$S family. In particular, $\pi$ ranks in a probabilistic sense the family of equilibria within and across different $k$s (see theorem \ref{thm:equlibriumranking}). This is particularly relevant for implementing the model when $\pi$ can be treated as the likelihood. In addition, the expression in \eqref{logl} provides for a transparent identification of model's parameters. It is clear that, given the variation in the data of individual choices $\{a_i\}_{i=1}^n$, friendships $\{ g_{ij} \}_{i,j=1}^n$ and attributes $\{ X_i \}_{i=1}^n$, functional forms for $v,w,q,\psi,\phi,\phi_S,\phi_N$ will be identified as long as the different parameters induce different likelihoods of the data. Finally, a closed-form expression for $\pi$ facilitates the use of likelihood-based methods for estimating model's parameters. 

\subsection{Speed of convergence}
The $k$CDs with random utility depend on $k$ in an important way despite the fact that their stationary distribution is invariant to $k$. The next result studies this dependence in isolation from all other determinants of the $k$CDs with random utility. 

\begin{theorem}\label{thm:convergencespeed}
\textsc{[$k$CDs ranking]}
Set $v_i=w_{ij}=h=\phi=\phi_S=\phi_N=q=\psi=0$. Then, the second eigen value of the $2^{(n^2+n)/2}$-by-$2^{(n^2+n)/2}$ transition matrix of the  $k$CD is given by:
\begin{equation}\label{eigen2}
\lambda_{k,[2]}=\frac{1}{n}\left(n-1+\frac{n-k}{n-1}\right)
\end{equation}
In particular, $\lambda_{k',[2]}<\lambda_{k,[2]}$ for $2 \leq k < k' \leq n$ so that the $k'$CD converges strictly faster than $k$CD to the stationary distribution $\pi$.
\end{theorem}

In the hypothesis of theorem \ref{thm:convergencespeed}, all payoff parameters in equations (\ref{payoff}-\ref{payoff3}) are set to zero so that players do not differentiate between different networks (i.e. $u_i(S;X)=0$ for all $S\in \Gn$) which implies that the $k$CDs traverse in unbiased way the space of all possible networks $\Gn$. In the end, the stationary distribution $\pi$ is one where the behaviors and network links are i.i.d. $\mathrm{Poisson(0.5)}$ and, importantly, $k$ is the only determinant of $k$CDs' transition probabilities and convergence rates.%
\footnote{In general, the shape of the potential, i.e. the terms of the potential function, and the geography of the network will likely influence the speed of convergence. To the best of my knowledge, treatment of the general case remains out of reach.}

There are two rationales behind pursuing a characterization of the speed of convergence of $k$CDs. As anticipated (and formally established shortly) $\pi$ probabilistically ranks the family of NE$k$S networks. In a dual fashion, the differential speed of convergence provides a means to rank the family of $k$CDs with random utilities. In particular, the larger $k$ is, the smaller is the second eigen value $\lambda_{k,[2]}$, i.e. the faster $k$CDs converge to $\pi$ (see \citealp[Section 4]{DebreuHerstein1953}). In this sense, a snapshot of the state (drawn from $\pi$) is more likely to reflect a $k$CD with random utility where $k$ is large as opposed to one where $k$ is small.

The second reason for why properties of $k$CDs are of their own interest is highlighted by \cite{BhamidiBrestlerSly_2011} who show that adaptive dynamic with local updates (i.e. $o(n)$ links at a time) converges very slowly. Such slow convergence rates could question the conceptual treatment of the limiting distribution $\pi$ as a likelihood. For this same reason, simulation based methods that rely on local updates may not work in practice for estimation/simulation of these models.%
\footnote{See the discussion in \cite{ArunJackson2016}.} 
Note that $k$CDs encompass not only local updates, e.g. $k=[n/2]$, and thus suggest a way to avoid the problem of slow convergence (poor approximation). Relatedly, theorem \ref{thm:convergencespeed} offers insights into an important trade-off for sampling design: the Markov chain is facing a trade-off between speed of convergence and complexity in simulating the next step. For small $k$, the convergence to $\pi$ is slower, however, the update is drawn from a discrete distribution with small ($2^k$) support. The opposite holds when $k$ is large.\footnote{The structure of the problem permits a substantial computational shortcut within the MH algorithm for generating the update of $k$CD. In particular, for any $k$ computing the acceptance probability scales only quadraticly with the size of the network because it is enough to compute the change in potential as opposed to the potential itself. The published code of the paper contains more details.}

\subsection{Discussion}

\subsubsection{Probabilistic ranking. The most probable equilibria}

The stationary distribution obtained in theorem \ref{thm:stationary} gives an intuitive (probabilistic) ranking of the family of NE$k$S networks. Under $\pi$, a network state will receive a positive probability, although it may not be an equilibrium in any sense. It will be desirable, however, that in the vicinity of an equilibrium, the equilibrium to receive the highest probability. Relatedly, the mode of $\pi$ (i.e. the state with the highest probability) has special role. This offers a new perspective to the theoretical results on equilibrium selection from evolutionary game theory, namely equilibrium ranking.

To formalize our discussion, define the neighborhood $\Ns \subset \Gn$ of $S \in \Gn$ as:
\begin{equation*}
  \Ns(S) = \left\lbrace (g_{ij},S_{-ij}): i\neq j,g_{ij}\in\{0,1\} \right\rbrace \bigcup
           \left\lbrace (a_{i},S_{-i}   : a_i \in \{0,1\} \right\rbrace
\end{equation*}

\begin{theorem} \label{thm:equlibriumranking}
Suppose assumptions \ref{asm:potential}-\ref{asm:meet2} hold.
\begin{enumerate}
  \item A state $S\in \Sn$ is a Nash equilibrium in a pair-wise stable network iff if it receives the highest probability in its neighborhood $\Ns$.
  \item The most likely network states $\Smode \in \Sn$ (the ones where the network spends most of its time) are pairwise Nash networks.
\end{enumerate}
\end{theorem}

\subsubsection{A $k$CD with random $k$}

Consider what appears to be a very unrestrictive meeting process, where every period a random individual meets a set of potential friends of random size and composition. Let $\kappa$ be a discrete process with support $2,\ldots,n$ and augment the meeting process with an additional initialization step with respect to the dimension of $\mu$. In particular, at each period first $\kappa$ is realized and then $\mu^k$ is drawn just as before. It is relatively straightforward to establish, without any assumptions on the process $\kappa$, that this augmented process has the same stationary distribution $\pi$ as the one from theorem \ref{thm:stationary}.\footnote{A formal statement and a proof are omitted because these follow the ones of theorem \ref{thm:stationary}.} This is another demonstration of the fact that different meeting processes result in observationally equivalent models.

\section{Data and estimation}

\subsection{The Add Health data}
The National Longitudinal Study of Adolescent Health is a longitudinal study of a sample of adolescents in grades $7$--$12$ in the United States in the $1994$--$95$ school year. The sample is representative of US schools with respect to region of country, urbanicity, school size, school type, and ethnicity. In total, 80 high schools were selected together with their ``feeder`` schools. The students were first surveyed in-school and then at home in four follow-up waves conducted in $1994$--$95$, $1996$, $2001$--$02$, and $2007$--$08$. This paper makes use of Wave I of the in-home interviews with students enrolled in the schools from the so called saturated sample. Only for schools from the saturated sample, all of their students were eligible for in-home interviews.

The in-home interviews contain rich data on students' behaviors, home environment, and friendship networks. These data are merged with administrative data on the average price of a carton of cigarettes from the American Chamber of Commerce Research Association (ACCRA). ACCRA's data are linked to the Add Health data on the basis of state and county FIPS codes for the year in which the data were collected. Additional details about the estimation sample including sample construction and sample statistics are presented in the appendix.

\subsection{Bayesian estimation}
The $k$CDs with random utility deliver a unique stationary distribution $\pi$ which for estimation purposes can be thought of as likelihood. Because no information is available on when the process started or on its initial state, the best prediction about the current state is given by $\pi$. For a single observation $S \in \Sn$, the likelihood is given by:%
\begin{equation}
  p(S|\theta) = \frac{\exp\{\Pfn_{\theta}(S)\}}{H_{\theta}}
\end{equation}
where $\Pfn_\theta$ is the potential (evaluated at $\theta$) and $H_{\theta}=\sum_{S \in \Sn} \exp\{S\}$ is an (intractable) normalizing constant.\footnote{The summation in calculating $H_\theta$ cannot be computed directly for practical purposes even for small $n$, e.g., for $n=10$ this summation includes $2^{55}$ terms.} The specific form of the likelihood pertains to the exponential family, whose application to graphical models has been termed as Exponential Random Graph Models (ERGM).\footnote{ERGMs are a broad class of statistical models, capable of incorporating arbitrary dependencies among the links of a network. See \cite{FrankStrauss1986markov} and \cite{WassermanPattison1996}.} 

The estimation draws from the Bayesian literature on approximating likelihoods with intractable normalizing constant developed in \cite{murray2012mcmc} and \cite{Liang2010double}. The proposed implementation augments their algorithm with an extra step informed by properties of the $k$CDs in the proposed model. 

The posterior sampling algorithm is exhibited in table \ref{table:algorithm}. In the original double M-H algorithm, an M-H sampling of $S$ from $\pi_\theta(S)$ is nested in an M-H sampling of $\theta$ from the posterior $p(\theta|S)$. The new piece in table \ref{table:algorithm} is the random meeting process in step $5$. Theorem \ref{thm:convergencespeed} suggests that varying $k$ improves the convergence and theorem \ref{thm:stationary} demonstrates that changing $k$ leaves the stationary distribution unaltered. Proposition \ref{prop:varying_mh} below demonstrates the validity of the algorithm.

\begin{table}[!t]
\caption{Varying double M-H algorithm}
\label{table:algorithm}
\begin{center}
Input: initial $\theta^{(0)}$, number of iterations $T$, size of the Monte Carlo $R$, data S \\
\begin{tabular}{cl}
\hline
1. & \textbf{for} $t=1\ldots T$ \\
2. & \quad Propose $\theta' \sim q(\theta';\theta^{(t-1)},S)$\\
3. & \quad Initialize $S^{(0)}=S$\\
4. & \quad \textbf{for} $r=1\ldots R$ \\
5. & \quad \quad Draw $k \sim p_k(k)$\\
6. & \quad \quad Draw a meeting $\mu(i,I_k)$ where $i\in \{1\ldots N\}$ and $I_k \subset \{1\ldots N\}\backslash\{i\}$ from $q_\mu(i,I_k)$\\
8. & \quad \quad Propose $S'$ where $(a_i,\{g_{ij}\}_{j\in I_k})$ are drawn from $q_\mu(S'|S^{(r-1)};(i,I_k))$ \\
9. & \quad \quad Compute 
$\bar{a} = \frac{\exp\{\Pfn_{\theta'}(S')\}}{\exp\{\Pfn_{\theta'}(S^{(r-1)})\}} 
           \frac{Q(S^{(r-1)}|S';p_k,q_{i,I_k})}{Q(S'|S^{(r-1)};p_k,q_{i,I_k})}$ \\
10. & \quad \quad Draw $a\sim \text{Uniform}[0,1]$\\
11. & \quad \quad If $a<\bar{a}$ then $S^{(r)}=S'$ else $S^{(r)}=S^{(r-1)}$\\
12. & \quad \textbf{end} for $[r]$ \\
13. & \quad Compute 
$\bar{a}=\frac{q(\theta^{(t-1)};\theta')}{q(\theta';\theta^{(t-1)})}
\frac{p(\theta')}{p(\theta^{(t-1)})}
\frac{\exp\{\Pfn_{\theta^{(t-1)}}(S^{(R)})\}}{\exp\{\Pfn_{\theta^{(t-1)}}(S)\}}
\frac{\exp\{\Pfn_{\theta'}(S)\}}{\exp\{\Pfn_{\theta'}(S^{(R)})\}}$\\
14. & \quad Draw $a\sim \text{Uniform}[0,1]$\\
15. & \quad If $a<\bar{a}$ then $\theta^{(t)}=\theta'$ else $\theta^{(t)}=\theta^{(t-1)}$\\
16. & \textbf{end} for $[t]$ \\ \hline
\end{tabular}
\end{center}
\end{table}

\begin{proposition}
\textsc{[Varying double M-H algorithm]}
\label{prop:varying_mh}
Let $1<k\le n$ and suppose assumptions \ref{asm:meet1} and \ref{asm:error_utility} hold. If in the algorithm of table \ref{table:algorithm}, the proposal density conditional on meeting $(i,I_k)$, $q_\mu(S'|S);(i,I_k))$ is symmetric, then the unconditional proposal $Q(S'|S)$ is symmetric. In particular, the acceptance ratio of the inner M-H step 9 does not depend neither on $p_k$ and nor on $q_\mu$.
\end{proposition}

The Bayesian estimator requires specifying prior distributions and proposal densities. All priors $p(\theta)$ are normal and all proposals ($p_k$, $\mu$, and $q_\mu$) are uniform over their respective domains.

\subsection{Parametrization}
The payoffs from \eqref{payoff} and \eqref{payoff2} have six sets of parameters: $v_i$, $w_{ij}$, $q$, $\phi$, $\phi_S$ and $\phi_N$. In the empirical specification, the first three are functions of the data $v_i=V(X_i)$, $w_{ij}=W(X_i,X_j)$, $q_{ijk}=q(X_i,X_j,X_k)$. 
Careful scrutiny of the data and extensive experimentation with various parametrizations motivate the final specification which is discussed in the appendix (appendix \ref{app:parameters} on page \pageref{app:parameters}). 

\subsection{Identification}
Because the model pertains to the exponential family, identification within the framework of many networks follows immediately. Indeed, a corollary of theorem \ref{thm:stationary} is that the likelihood of the model is proportional to $\exp\left\{\sum_{r=1}^R\theta_i w_i(S,X) \right\} $, where $w_i:\Sn \times \Xn \longrightarrow \R$ are functions of the data. To obtain identification, it is enough that the sufficient statistics $w_i$ are linearly independent functions on $\Sn \times \Xn$ (e.g., see \cite{lehmann_casella_98} for a textbook treatment). In the structural model above, this condition is readily established.\footnote{Most of the parameters are identified in the asymptotic frame where the size of the network grows to infinity (as opposed to the number of networks going to infinity). For example, turning off the externalities ($\phi=0$, $\phi_S=0$, $\phi_N=0$, $q=0$, $\psi=0$) implies that both smoking and friendships are independently distributed so that standard LLNs apply in the single large network asymptotics.}

\subsubsection*{Unobservable heterogeneity in friendship selection and decision to smoke}

In addition to the models' parameters for observable attributes, it is possible to incorporate agents' specific unobservable types $\tau_i \sim N(0,\sigma^2_{\tau})$ which may influence \emph{both} the utility for friendships, e.g. $W(.,.)$ could include term $|\tau_i-\tau_j|$, and also the propensity to smoke, e.g. $V(.)$ could include a term $\rho_{\tau}\tau_i$. In this case the likelihood has to integrate out $\vec{\tau}$:
\begin{equation}
p(S|\theta) = \int_{\vec{\tau}} \frac{\exp\{\Pfn_\theta(S,\vec{\tau}\}}{\sum_{\hat{S}} \exp\{\Pfn_\theta(\hat{S},\vec{\tau})\} } \phi(\vec{\tau}) d\vec{\tau}
\end{equation}

There are a couple of approaches to discuss identification in this case. Within the Bayesian paradigm, identification casually obtains as long as the data provides information about the parameters. Even a weakly informative prior can introduce curvature into the posterior density surface that facilitates numerical maximization and the use of MCMC methods. However, the prior distribution is not updated in directions of the parameter space in which the likelihood function is flat (see \citealp{AnSchorheide2007bayesian}). From a frequentist perspective, the heuristic identification argument goes as follows. Friends who are far away in observables, must have realizations of the unobservables very close by. If in the data those individuals are either smokers or non smokers with very high probability then it must be the case that $\rho_\tau$ is large. However, formalizing this argument is nether immediate nor it is clear whether this argument will support non-parametric identification so this endeavor is left for future research.


\subsection{Estimation results}

\begin{table}[t!]
\begin{center}
\caption{Parameter estimates (posterior means)} \label{tab:estimates}
\hspace*{-1cm}
\scalebox{0.8}{
\begin{tabular}{llcccccc}
\hline \hline
\multicolumn{4}{c}{\textit{Utility of smoking}}  \\
 & Parameter   & No net data & Exog net & No PE & Model \\ \hline  1 &Baseline probability of smoking          &  $0.12^{***}$ &  $0.17^{***}$ &  $0.21^{***}$ &  $0.18^{***}$ \\ 
  2 &Price $\times 100$                       &    $-0.17^{}$ &    $-0.21^{}$ & $-0.61^{***}$ &   $-0.24^{*}$ \\ 
  3 &Mom edu (HS\&CO)$^{MP}$                  & $-0.04^{***}$ & $-0.05^{***}$ & $-0.05^{***}$ & $-0.05^{***}$ \\ 
  4 &HH smokes                                &  $0.11^{***}$ &  $0.13^{***}$ &  $0.16^{***}$ &  $0.14^{***}$ \\ 
  5 &Grade 9+$^{MP}$                          &  $0.18^{***}$ &  $0.16^{***}$ &  $0.24^{***}$ &  $0.16^{***}$ \\ 
  6 &Blacks$^{MP}$                            &  $-0.3^{***}$ &  $-0.3^{***}$ & $-0.35^{***}$ & $-0.31^{***}$ \\ 
  7 &$30\%$ of the school smokes$^{MP}$       &  $0.07^{***}$ &  $0.05^{***}$ &            -- &  $0.05^{***}$ \\ 

\\
\multicolumn{4}{c}{\textit{Utility of friendships}} \\
& Parameter   & No net data & Exog net & No PE & Model \\ \hline  8 &Baseline number of friends               &            -- &            -- &  $4.63^{***}$ &   $3.4^{***}$ \\ 
  9 &Different sex$^{MP\%}$                   &            -- &            -- & $-0.72^{***}$ & $-0.72^{***}$ \\ 
 10 &Different grades$^{MP\%}$                &            -- &            -- & $-0.89^{***}$ & $-0.89^{***}$ \\ 
 11 &Different race$^{MP\%}$                  &            -- &            -- &   $-0.33^{*}$ & $-0.39^{***}$ \\ 
 12 &Cost/Economy of scale                    &            -- &            -- & $-0.21^{***}$ & $-0.22^{***}$ \\ 
 13 &Triangles$^{MP\%}$                       &            -- &            -- &  $1.18^{***}$ &  $1.22^{***}$ \\ 
 14 &$\phi_{smoke}^{MP}$                      &            -- &  $0.04^{***}$ &            -- &  $0.05^{***}$ \\ 
 15 &$\phi_{nosmoke}^{MP}$                    &            -- &  $0.03^{***}$ &            -- &  $0.04^{***}$ \\ 

\\ \hline
\end{tabular}%
} 
\end{center}
\fignotetitle{Note:} \fignotetext{MP stands for the estimated marginal probability in percentage points and MP$\%$ for estimated marginal probability in percent, relative to the baseline probability. The posterior sample contains $10^5$ simulations before discarding the first $20\%$. The shortest $90/95/99\%$ credible sets not including zero is indicated by $^{*}$/$^{**}$/$^{***}$ respectively.}
\end{table}

\begin{figure}[!h]
	\begin{center}
		\caption{Posterior distribution for the price parameter}
		\label{fig:priceposteriors}
		\includegraphics[scale=0.7]{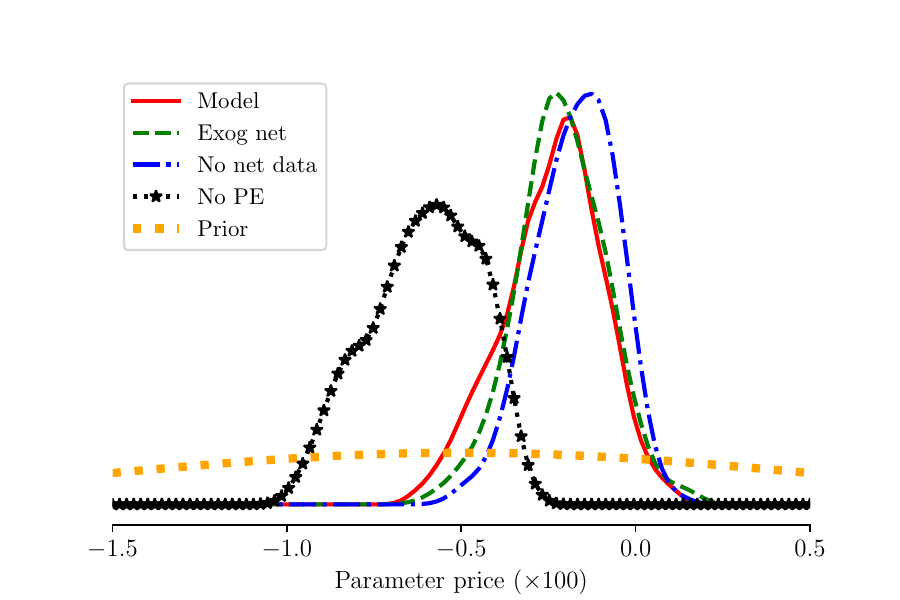}
	\end{center}
	\fignotetitle{Note:} \fignotetext{The hypotheses for equal means between the model's posterior and each of the other posteriors on the plot are rejected with $p<0.01$ by $t$-tests.}
\end{figure}

Table \ref{tab:estimates} presents model's estimates (the posterior means) for four different estimation scenarios: (i.) without network data, (ii.) with fixed network, (iii.) without peer effects, and (iv.) the full model. The estimates have been transformed for ease of interpretation to baseline probabilities, marginal probabilities ($MP$ in ppt) and relative marginal probabilities ($MP\%$ in pct)\footnote{For example, the baseline probability of smoking $\theta_1$ is derived from the intercept $v_0$ as $\frac{e^{v_0}}{1+e^{v_0}}$. Superscript MP stands for marginal probability and MP$\%$ stands for marginal probability in percentages with respect to the baseline probability of smoking $\frac{e^{v_0}}{1+e^{v_0}}$. Appendix \ref{app:parameters} on page \pageref{app:parameters} provides details.}.  
It is worth pointing out that the estimate for the price coefficient does not vary much in magnitude (but only in significance). The point estimates in table \ref{tab:estimates} together with the posterior distributions of this parameter in figure \ref{fig:priceposteriors} suggest that the largest biases arise when peer effects terms are omitted (column ``No PE'') or when the econometrician does not have data on the friendship network (column ``No Net Data'').\footnote{The hypotheses  of equal means between the model's posterior and each of the other posteriors in figure \ref{fig:priceposteriors}  are rejected with $p<0.01$ by $t$-tests.} Nevertheless, it is difficult to interpret the magnitudes of these differences nor the magnitudes of the structural estimates altogether in a concrete economic context. This is the case because the reported marginal effects are first order approximations which do not take into account the overall equilibrium response of the system.\footnote{\label{footnote:pe}A related point is that the parameter $\phi_S$ cannot be interpreted as the effect on the likelihood of smoking from a randomly assigned friend who is a smoker because, in the model, individuals cannot be forced into friendships. Rather, individual's utility increases with $\phi_S$ (or $\phi_N$) with every instance where her choice to smoke (or not) and her choice of a friend are such that she and this friend of hers both smoke (or not).}

\begin{figure}[!h]
	\begin{center}
		\caption{Posterior distribution for the local PE parameters}
		\label{fig:PEposteriors}
		\includegraphics[scale=0.7]{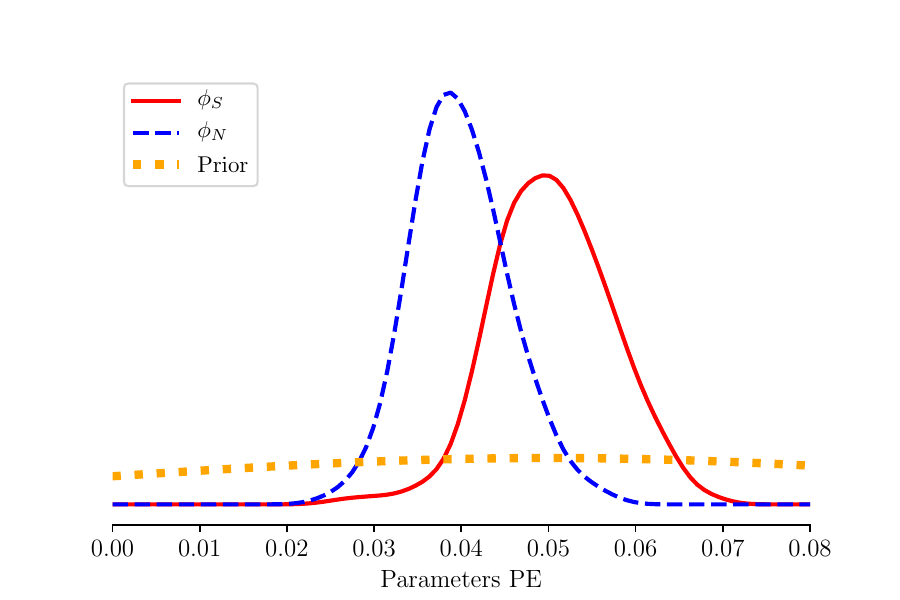}
	\end{center}
	\fignotetitle{Note:} \fignotetext{The hypotheses for equal means and equal distributions between the parameters for peer effects among smokers $\phi_S$ and among non-smokers $\phi_N$ are rejected with $p<0.01$.}
\end{figure}

A final point on the estimation results is that the peer effect externalities are very different between smokers compared to those between non-smokers. Figure \ref{fig:PEposteriors} reveals that the peer pressures between smokers is much stronger than that of non-smokers.(see footnote \ref{footnote:pe})

\newpage

\subsection{Model fit}

\begin{table}[htbp]
\caption{Model fit}
\label{table:fit}
\begin{center}
\begin{tabular}{lcc}
\hline \hline
\multicolumn{3}{c}{\textit{Selected moments}} \\
    Moment & Model & Data \\ \hline 
              Prevalence & 0.410 (0.408) & 0.408  \\ 
                 Density & 0.007 (0.005) & 0.005  \\ 
              Avg degree & 1.250 (0.966) & 0.973  \\ 
              Min degree & 0.275 (0.000) & 0.000  \\ 
              Max degree & 4.808 (4.568) & 5.308  \\ 
        $a_ig_{ij}a_j/n$ & 0.543 (0.253) & 0.256  \\ 
$(1-a_i)g_{ij}(1-a_j)/n$ & 0.400 (0.396) & 0.404  \\ 
           Two-paths$/n$ & 0.639 (0.490) & 0.501  \\ 
           Triangles$/n$ & 7.686 (0.023) & 0.066  \\ 

\\
\multicolumn{3}{c}{\textit{Mixing patterns}} \\ 
                      HI & 0.239 (0.231) & 0.236  \\ 
                     CHI & -0.300 (-0.299) & -0.303  \\ 
                     FSI & 0.665 (0.667) & 0.662  \\ 

\hline \\
\end{tabular}
\end{center}

\fignotetitle{Note:} 
\fignotetext{Columns Data and Model compare selected moments of the estimation sample with those of synthetic data 
generated by the estimated model. For the latter mean and median are reported (median in parentheses). 
Two-paths is defined as $\sum_{i>j} g_{ij}g_{il}(1-g_{il})$. Triangles is defined as $\sum_{i>j>l}g_{ij}g_{il}g_{il}$.
The Homophily index (HI), Coleman homophily index (CHI), and Freeman segregation index (FSI) 
are measures of the mixing patterns between students with the same smoking statuses (see also table \ref{table:fit_mixing}). 
For more details about computing those indices, see \cite{CurrariniJacksonPin2010} Definitions 1 and 2 in the supplemental appendix.}
\end{table} 


\begin{table}[htbp]
\caption{Fit mixing matrix (model left, data right)}
\label{table:fit_mixing}
\centering
\begin{footnotesize}
\begin{tabular}{llcccc}
  &  & \multicolumn{2}{c}{\axislabel{Nominee}}                                 & \multicolumn{2}{c}{\axislabel{Nominee}} \\
  \multirow{5}{*}{\rotatebox[origin=c]{90}{\axislabel{Nominator}}}
  &            & Smoker               & Nonsmoker                      & Smoker               & Nonsmoker           \\\cmidrule{2-6}
& Smoker   & \textbf{  65\% (56.6)} &   35\% (30.1) & \textbf{  63\% (52.1)} &   37\% (30.4)\\ 
 & Nonsmoker   &   29\% (30.1)& \textbf{   71\% (74.0)} &   29\% (30.4) & \textbf{   71\% (75.4)}
\\
\cmidrule{2-6}
\end{tabular}

\end{footnotesize}
\end{table}

Table \ref{table:fit} compares statistics from the data to statistics from a sample generated with the estimated model. This is a sample of size $1000$ where each draw is generated via a long-run ($20,000$ draws) of the $k$CD with random utility parametrized with a draw from the posterior. In addition to statistics that are directly targeted by the model's parameters (overall prevalence, density, and average degree), statistics which are only indirectly governed by model's parameters are reported in tables \ref{table:fit} and \ref{table:fit_mixing}, e.g. maximum degree, certain friendship configurations, mixing etc.

Overall the model fits well the smoking decisions and the network features of the data. The only caveat is the number of triangles as fraction of the size of the network which in the data is $0.066$ while the draws from the model are right-skewed (i.e., have a long tail to the right) with mean of $7.686$ and median of $0.023$. This is due to the presence of very few draws with very densely connected networks. The most likely reason for this discrepancy is that in the model triangles are generated only via a single parameter which does not depend on observables, i.e. race, sex etc. This parsimonious specification is dictated by the small sample size and further exploration of this feature is left for the future.

\section{Policy experiments}
\subsection{A. Changes in the price of tobacco}

The estimated model serves as a numerical prototype for the equilibrium behaviors and, in particular, for the equilibrium adjustments to various policy interventions. 
Table \ref{table:ctrf-price} presents simulated increases in tobacco prices ranging from $20$ to $160$ cents (in the sample tobacco prices average at $\$1.67$ for a pack) and their effect on the overall tobacco smoking rates for the sample. The table compares the predictions from the full model to those from the model when agents are restricted from adjusting their friendship links and those from a model that is estimated without data on the friendship network.


    \begin{table}[htbp]
    \caption{The effect on smoking rate from changes in the price of tobacco}
    \label{table:ctrf-price}
    \begin{center}
    \begin{tabular}{cccccc}
    Price increase & Model  &  Exog net & No net data\\ \hline \hline
      20 &  2.5 &  2.2 &  1.3 \\ 
  40 &  4.7 &  4.2 &  2.6 \\ 
  60 &  6.9 &  6.1 &  3.9 \\ 
  80 &  8.7 &  7.9 &  5.1 \\ 
 100 & 10.3 &  9.4 &  6.2 \\ 
 120 & 11.8 & 10.9 &  7.4 \\ 
 140 & 13.1 & 12.3 &  8.4 \\ 
 160 & 14.3 & 13.5 &  9.5 \\ 

    \hline
    \end{tabular}
    \end{center}
    \fignotetitle{Note:} The first column shows proposed increases in tobacco prices in cents. 
    The average price of a pack of cigarettes is \$1.67 so that 20 cents is approximately 10\%. 
    The second through fourth columns show the predicted increase in the overall smoking (baseline 41\%) in ppt 
    from the full model, from the model when the friendship network is fixed, and 
    from the model when no social network data is available (i.e., $\phi_S=\phi_N=0$). 
    \end{table}
    
\begin{figure}[htbp]
	\begin{center}
		\caption{Distribution of the effect on smoking rate from selected price changes}
		\label{fig:priceeffects}
		\includegraphics[scale=0.6]{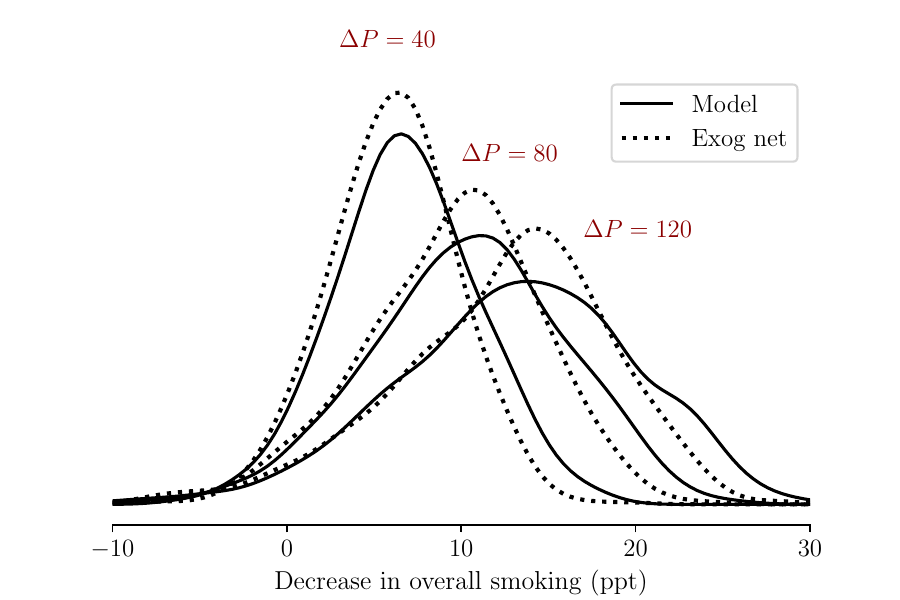}
	\end{center}
	\fignotetitle{Note:} In the model with exogenous friendships the distribution of the predicted smoking rates underestimates both the mean smoking rates (table \ref{table:ctrf-price}) and the amount of uncertainty associated with the policy intervention (i.e, the variance of the distribution) as compared with the full model. The appendix provides statistical tests for these differences.
\end{figure}

As seen in table \ref{table:ctrf-price}, smoking rates responds to price changes. Comparison between model's predictions with and without friendship adjustments (columns two and three) reveals that the latter underestimates the mean response by around $15\%$ . In addition, the model without friendship choices underestimates the variance of this response as well (see figure \ref{fig:priceeffects}). Finally, lack of network data (forcing the restriction $\phi_S=\phi_N=0$) leads to a bias in the mean response to price changes that is between $50\%$ and $70\%$ of the prediction of the full model.

This analysis suggests that the freedom of breaking friendships and changing smoking behavior induces slightly larger decrease in overall smoking compared to situation when individuals are held in their existing (fixed) social networks. Figuratively, a price change has two effects on the decision to smoke: the direct effect operates through changing individuals' exogenous decision environment and the indirect/spillover effect operates through changing the peer norm which then puts additional pressure on the individuals' to follow the change. When comparing the endogenous to fixed network, the direct effect is likely to be stronger in the former environment while the indirect effect is likely to be stronger in the latter environment.\footnote{It is interesting to relate this findings to the theoretical analysis in \cite{Jackson2016Paradox} who argues that variability in individuals' popularity (degree in a social network) leads to biased perceptions for the social norm which in turn leads to higher levels of activities compared to a situation when there is no variability in individuals' popularity. This counterfactual experiments hints to such amplification mechanisms (in quite different settings).}
Related to this decomposition, this study suggests that quantitatively the direct effect dominates in shaping the overall equilibrium adjustments.

\subsection{B. Changes in the racial composition of schools}

Suppose that in a given neighborhood there are two racially segregated schools: ``White School'' consisting of only white students and ``Black School'' consisting of only black students. One would expect that the smoking prevalence in White school is much higher compared to Black school because, in the sample, black high students smoke three times less than white high school students. Consider a policy aiming to promote racial desegregation, which prevents schools from enrolling more than $x$ percent of students of the same race. If such policy is in place, will students from different races form friendships and will these friendships systematically impact the overall smoking in one or another direction?

One of the racially balanced schools in the sample is used to evaluate the effect of this policy.\footnote{The school has 150 students of which $40\%$ are Whites and $42\%$ are Blacks. It incorporates students from grades 7 to 12. From these, the simulations use students from grades 10 to 12 because older students are more likely to form meaningful friendships and to smoke.} In particular, the Whites and the Blacks from this school serve as prototypes for the White School and Black School respectively.
To implement the proposed policy, a random set of students from the White School is swapped with a random set of students from the Black School. For example to simulate the effect of a $70\%$ cap on the same-race students in a school, a swap of $30\%$ is simulated. 


    \begin{table}[t!]
    \caption{The effect on smoking rates from same-race students caps}
    \label{table:ctrf-schoolcomposition}
    \begin{center}
    \begin{tabular}{cccc}
    Same-race & School & School & Overall \\
    cap ($\%$) & White  & Black   & \\ \hline \hline
      0 & 32.9 &  4.5 & 18.7 \\ 
 10 & 29.2 &  6.7 & 17.9 \\ 
 20 & 25.6 &  9.3 & 17.4 \\ 
 30 & 23.6 & 11.1 & 17.4 \\ 
 40 & 18.8 & 15.0 & 16.9 \\ 
 50 & 17.0 & 16.8 & 16.9 \\ 

    \hline
    \end{tabular}
    \end{center}
    \fignotetitle{Note:} \fignotetext{A cap of $x\%$ same-race students is implemented with a swap of $(100-x)\%$ students. 
    The last column shows the predicted changes in overall smoking under different same-race caps. 
    The policy induces statistically significant changes in the overall smoking as suggested by 
    the statistical tests in appendix D.}
    \end{table}

Table \ref{table:ctrf-schoolcomposition} presents the simulation results, which suggest that racial composition affects the overall smoking prevalence. The first column shows the size of the set of students which is being swapped. The second, third, and forth columns show the simulated smoking prevalence in the White School, Black School, and both, respectively. The table suggest that overall smoking prevalence is lower when schools are racially balanced, thus supporting policies promoting racial integration in the context of fighting high smoking rates.%
\footnote{The appendix demonstrates that these differences have statistical power.} 

It is important to note that the simulations here offer only suggestive evidence on the role of racial desegregation on the overall prevalence of smoking. There are many factors, e.g. the profile of all observables for the entire schools (income, home environment, tobacco price,  etc), that are likely to influence the outcome of desegregation. Unfortunately, the Add Health data does not offer substantial variation in those factors and the empirical analysis relies on a (the only) racially balanced school in the data. The author hopes this study to stimulate further research into this question. 

\subsection{C. Aggregate effects of an anti-smoking campaign}
The last experiment considers the effects of an anti-smoking campaign that can prevent with certainty a given number of students from smoking. An example of such intervention is a weekend-long information camp on the health consequences of smoking. Assuming that the camp is very effective in terms of preventing students from smoking but it is too costly to enroll all students in this camp, the question is once the ``treated students'' come back will their smoking friends follow their example and stop smoking, or will their friends un-friend them and continue smoking?


    \begin{table}[!h]
    \caption{Spillovers}
    \label{table:ctrf-spillovers}
    \begin{center}
    \begin{tabular}{ccccc}
    \multirow{2}{*}{Campaign (\%)}  & \multirow{2}{*}{Smoking}  & \multicolumn{1}{c}{Predicted effect} & Actual & \multirow{2}{*}{Multiplier} \\
    & & proportional & effect \\
     \hline \hline
    - & 42.1 & - & - & \\ 
   3& 39.6&  1.3&  2.6&  2.0 \\ 
   5& 38.2&  2.1&  3.9&  1.9 \\ 
  10& 34.6&  4.2&  7.5&  1.8 \\ 
  20& 28.7&  8.4& 13.4&  1.6 \\ 
  30& 23.5& 12.6& 18.6&  1.5 \\ 
  50& 15.1& 21.1& 27.0&  1.3 \\ 

    \hline
    \end{tabular}
    \end{center}
    \fignotetitle{Note:} The first column lists the alternative attendance rates. The second and third columns display the smoking rate and the change in smoking rate respectively if the decrease would be proportional to the intervention, i.e. computes a baseline without peer effects. The last column computes the ratio between the percentage change in the number of smokers and the attendance rate. Note that that attendance is random with respect to the smoking status of the students. If the campaign is able to target only students who are currently smokers, the spillover effects will be even larger.
    \end{table}

Table \ref{table:ctrf-spillovers} presents the simulation results with two schools that feature smoking rates at the sample mean. The table suggests that an anti-smoking campaign may have a large impact on the overall prevalence of smoking, without necessarily being able to directly engage a large part of the student population.\footnote{The policy is simulated $10^3$ times, where each time a new random draw of attendees is being considered.} In particular, the multiplier factor--the ratio between the actual effect and effect constrained to the treated sub-population--indicated a substantial spillover effects reaching up to the factor of 2. These spillover effects operate through the social network, from those who attended the camp to the rest of the school.

\section{Concluding remarks}
The premise of this paper is that individuals may respond differently to changes, with some following their friends' behaviors and others breaking away from their old friends in a search for new friends that will accept their new behaviors. This decision environment involves fundamentally different choices and generates complex mathematical structures. In equilibrium, players internalize the need for consensus in forming friendships and choose their optimal strategies on subsets of $k$ players - a form of bounded rationality. The $k$-player consensual dynamic delivers a probabilistic ranking of the proposed equilibria, and, via a varying $k$, facilitates the implementation of the model.

The estimation of a structural model of adolescents' smoking and friendships demonstrates that peer effect complementarities between smokers are substantially stronger than those between non-smokers. It also documents the estimation biases due to not accounting for the endogeneity of the friendship network and those due to the lack of social network data. Counterfactual analysis with the estimated model suggests that:
(a.) the response of the friendship network to changes in tobacco price amplifies the intended effect of price changes on smoking, 
(b.) racial desegregation of high-schools decreases the overall smoking prevalence, 
(c.) the peer effect complementarities are substantially stronger between smokers compared to between non-smokers,
(d.) the magnitude of the spillover effects from small scale policies targeting individuals' smoking choices are roughly double compared to the scale of these policies.

Overall this paper formulates an avenue to study the complementarities and coordination in live social networks, i.e. social networks that adapt to the behaviors of individuals. The literature has just started to understand the forces present in these environments (e.g., see \cite{Jackson2016Paradox}) while the empirical investigation of many hypothesis remains for the future (e.g., \cite{CarrellSacerdoteWest2013}, \cite{graham2014complementarity}).

\newpage
\appendix
\section{Proofs}
\label{section:proofs}

\begin{proof}[Proposition \ref{prop:neksn-exist}(on p. \pageref{prop:neksn-exist})] \label{proof:prop-neksn-exist}
Note that $\Delta_{g_{ij}}u_i() = \Delta_{g_{ij}}u_j()$. This property of the preferences implies that the unconstrained maximum in \eqref{iproblem} is feasible w.r.t. the stability constraints \eqref{stability_constraint}. That is, for any $i$ and $I_k = \{i\}\cup\{i_1,\ldots,i_{k-1}\}$ the solution of individual's decision problem (\ref{iproblem}-\ref{stability_constraint}) is simply
\begin{equation}\label{iproblem-representation} \tag{\ref{iproblem}}
\mathop{\mathrm{argmax}}_{\substack{a_i,g_{ij}\\ j\in \Ik}} \Pfn(S).
\end{equation}
This completes the proof of part one because \eqref{iproblem-representation} always has a solution. 

For part two, the first step is to extend the property $\Delta_{g_{ij}}u_i() = \Delta_{g_{ij}}u_j()$ to a deeper property of the preferences namely that the preferences of all players can be expressed by a single potential function.%
\footnote{The existence of potential implies $\Delta_{g_{ij}}u_i() = \Delta_{g_{ij}}u_j()$ but the converse is not true.} Indeed, consider $\Pfn : \Gn \times \mathbf{X}_n \longrightarrow \mathbb{R}$:
\begin{eqnarray} \label{pfn}
\Pfn(S,X)
&=& \sum_{i} a_{i} v(X_{i}) + \frac{1}{2}\sum_{i,j}g_{ij}w\left(X_{i},X_{j}\right) \\
& & +\frac{1}{2} \phi \sum_{i,j; i\neq j}a_{i}a_{j} 
    +\frac{1}{2}\phi_S \sum_{i,j}g_{ij}a_{i}a_{j}
    +\frac{1}{2}\phi_N \sum_{i,j}g_{ij}(1-a_{i})(1-a_{j}) \\
& & + \frac{1}{6}\sum_{i,j,k}q(X_i,X_j,X_k)g_{ij}g_{jk}g_{ki}
\end{eqnarray}
where $i\neq j$ is dropped from the summation ranges where possible because the convention that $g_{ii}$ is defined and equals to $0$ for all $i$ so that $\sum_{i,j;i\neq j} g_{ij}=\sum_{i,j} g_{ij}$. To show that $\Pfn$ is potential, it is sufficient to verify that (using assumption \ref{asm:potential}):
\begin{eqnarray}
\Delta_{a_i}u_i()      &= & v_i + \phi \sum_{j\neq i} a_j 
+ \phi_S \sum_{j\neq i} g_{ij} a_j - \phi_N \sum_{j\neq i} g_{ij} (1-a_j) \\ \notag
 &= & \Delta_{a_i}\Pfn() \\
\Delta_{g_{ij}}u_i()   &= & w_{ij} + q(X_i,X_j,X_k)\sum_k g_{ik}g_{jk} \\
& & - \psi (d_i+d_j) 
 + \phi_S a_i a_j + \phi_N(1-a_i)(1- a_j)\\ \notag
 &= & \Delta_{g_{ij}}\Pfn()
\end{eqnarray}

Next, fix $k$ and consider the following adaptive dynamic on $\Sn$. Every period draw at random $i$ and $I_k$ (from the uniform distributions over their respective domains), and let $i$ choose in her argmax \eqref{iproblem-representation}. For this dynamic, the value of the potential is nondecreasing so, invoking submartingale convergence argument, the potential convergences. Unless two states have the same potential (generically false), this implies that the state converges to a particular network which is, of course, a NE$k$S network. This same technology appears in the proof of proposition \ref{prop:dynamics}.
\qed
\end{proof}

As it will be useful later on, proposition \ref{neksn-representation} states characterization \eqref{iproblem-representation} in both directions. The proof of the if direction follows closely that of the only if direction above, and is omitted. 

\begin{proposition}\label{neksn-representation}
$S^*$ is a NE$k$S network \emph{iff} $\forall i, I_k = \{i\}\cup\{i_1,\ldots,i_{k-1}\}$
\begin{equation*}
(a^*_i,g^*_{ij})_{j\in \Ik} \in \mathop{\mathrm{argmax}}_{\substack{a_i,g_{ij}\\ j\in \Ik}} \Pfn((a_i,g_{ij})_{j\in \Ik};S^*_{-(a_i,g_{ij})_{j\in \Ik}})
\end{equation*}
\end{proposition}

\begin{proof}[Proposition \ref{prop:neksn-prop}(on p. \pageref{prop:neksn-prop})] \label{proof:prop-neksn-prop}
For $k=2$, definition \ref{def:eqlb} directly implies that a NE$k$S network is pairwise stable. Note that this observation is independent of the particular payoff structure here. 

Let $k=n$. That a NE$k$S network is pairwise stable follows from part 3 of this proposition (demonstrated next). To see that a NE$k$S network $S^*$ is a Nash network, consider the following strategies in a normal form link-announcement game (given the equilibrium behavior $\vec{a}^*$): each player announces his NE$k$S links. Proceeding by contradiction, for if a player has a profitable deviation then it would be possible to construct (appending $a_i^*$) an $S_{(i)}$ which she prefers to her NE$k$S play $S^*_{(i)}$. Therefore $S^*_{(i)} \notin \mathop{\mathrm{argmax}}_{\substack{a_i,g_{ij}\\ j\in \Ik}} u_i(S)=\mathop{\mathrm{argmax}}_{\substack{a_i,g_{ij}\\ j\in \Ik}} \Pfn(S)$ which contradicts proposition \ref{neksn-representation}.

Finally, the characterization from proposition \ref{neksn-representation} directly implies part three. In particular, if $k'<k$, $I_{k'} \subset I_k$ and $(a^*_i,g^*_{ij})_{j\in \Ik} \in \mathop{\mathrm{argmax}}_{\substack{a_i,g_{ij}\\ j\in \Ik}} \Pfn((a_i,g_{ij})_{j\in \Ik};S^*_{-(a_i,g_{ij})_{j\in \Ik}})$ then
$(a^*_i,g^*_{ij})_{j\in I_{k'}} \in
\mathop{\mathrm{argmax}}_{\substack{a_i,g_{ij}\\ j\in I_{k'}}} \Pfn((a_i,g_{ij})_{j\in I_{k'}};S^*_{-(a_i,g_{ij})_{j\in I_{k'}}}).
$
\qed
\end{proof}

\begin{proof}[Proposition \ref{prop:dynamics} (on p. \pageref{prop:dynamics}] \label{proof:prop-dynamics}
That any NE$k$S network is absorbing for the $k$CD follows from definition \ref{def:eqlb}. The second part follows from observing that $\Pfn_t$ is a submartingale, i.e., $E[\Pfn_{t+1}|S_t ] \geq \Pfn_t.$, so that $\{\Pfn_t\}$ converges almost surely. Because the network size is finite it follows that $\{\Pfn_t\}$ is constant for large $t$ and, generically, the same holds for $S_t$, i.e. $S_t=S^*$ for large enough $t$. Because of assumption \ref{asm:meet1} (any meeting is possible), this can happen only if $S^*$ is a NE$k$S network.
\qed
\end{proof}

\begin{proof}[Theorem \ref{thm:stationary} (p. \pageref{thm:stationary})] \label{proof:thm-stationary}
The first part follows from standard results on convergence of Markov chains. In particular, $k$-CDs with random utility induce a finite state Markov chain which, with assumptions \ref{asm:meet1} and \ref{asm:error_utility}, is irreducible, positive recurrent, and aperiodic. This is sufficient to obtain the conclusion of part one.

For the second part, it is enough to show that 
\begin{equation}\label{detailed_balance}
\Pr(S'|S;k)\exp\{\Pfn(S)\} = \Pr(S|S';k)\exp\{\Pfn(S')\},
\end{equation}
where $\Pr(S'|S;k)$ is the one step transition probability for moving from $S$ to $S'$.


There are two cases to consider: $\Pr(S'|S;k)=0$ and $\Pr(S'|S;k)>0$. Note that the hypothesis guarantees that $\Pr(S',S;k)>0$ iff $\Pr(S,S';k)>0$. Thus, if $\Pr(S'|S;k)=0$ then $\Pr(S|S';k)=0$ and, trivially, \eqref{detailed_balance} holds.

Consider the case $\Pr(S'|S;k)>0$. For fixed $k$, $S$, and $S$ let $\MK_{S'|S;k}$ be the set of all possible meetings that can result in transitioning from $S$ to $S'$. Note that for some triples ($S,S',k$), $\MK_{S'|S;k}$ is empty. However, if $\Pr(S'|S;k)>0$ then $\MK_{S'|S;k}\neq \emptyset$.

Let us pause with an example of this notation. Given the triple ($S,S',k$)
\begin{equation*}
\Pr(S'|S;k)=\sum_{\mu \in \MK_{S'|S;k}} \Pr(\mu) \frac{\exp\{u_{i}(S')\}}{\sum_{\hat{S} \in \NS_k(\mu,S)}\exp\{u_{i}(\hat{S})\}}.
\end{equation*}
Consider the case when $S$ and $S'$ agree on all $\{g_{ij}\}_{i\neq j}$ but differ in $a_{i}$ for some $i$, say $S=(a_{i}=0,S_{-a_i})$ and $S'=(a'_{i}=1,S_{-a_i})$. Then, $\MK_{S'|S;k}$ is the set of all possible meeting tuples $(i,I_{k-1})$ where player $i$ meets different $\{i_1,\ldots, i_{k-1}\}$, and the size of $\MK_{S'|S;k}$ is ${n-1 \choose k-1}$. To close the example, assume that all meetings are equally likely and that individuals are indifferent to all outcomes (i.e. $u_i$ is a constant). 
Then $\Pr(\mu)=\frac{1}{n}\frac{1}{{n-1 \choose k-1}}$ and $\frac{\exp\{u_{i}(S')\}}{\sum_{\hat{S} \in \NS_k(\mu,S)}\exp\{u_{i}(\hat{S})\}}=\frac{1}{2^k}$ so that
\begin{equation*}
\Pr(S'|S;k)={n-1 \choose k-1} \frac{1}{n{n-1 \choose k-1}} \frac{1}{2^k}=\frac{1}{n 2^k}.
\end{equation*}

Recall that ${\NS}_k(S,\mu)\subset \SSn$ denotes the set of all possible states that can result from the meeting $\mu$ following a state $S$. The proof follows from the following observations:\footnote{The proof of lemma \ref{neibourhoods} involves basic reasoning and is omitted. The challenging part is to state and interpret the lemma. Formal proof available upon request.}

\begin{lemma}
\label{neibourhoods}
For all $k$, $S$, $S'$, and $\mu=(i,I_{k-1})$:
\begin{enumerate} [topsep=0ex,parsep=0ex,itemsep=0ex,leftmargin=0.7cm,label=(\roman*)]
\item $\MK_{S'|S;k}=\MK_{S|S';k}$ for all $S,S'\in \SSn$;
\item $S' \in \NS_k(\mu,S)$ iff $S \in \NS_k(\mu,S')$;
\item If $S' \in \NS_k(\mu,S)$ then $\NS_k(\mu,S)=\NS_k(\mu,S')$.
\end{enumerate}
\end{lemma}

Part $(i)$ asserts that each meeting that can result in transitioning from $S$ to $S'$ may result in transitioning from $S'$ to $S$ as well (provided the starting state were $S'$). Part $(ii)$ re-states this observation in terms of the neighborhoods of $S$ and $S'$ given a meeting $\mu$. Finally, part $(iii)$ notes that if a meeting $\mu$ could result in $S$ transiting to $S'$, then the set of all feasible states following $\mu$ and $S$ coincides with the set of all feasible states following $\mu$ and $S'$.

From lemma \ref{neibourhoods}, the one step transition probability can be written as:
\begin{eqnarray}
\Pfn(S) \Pr(S'|S;k) & = &\Pfn(S) \sum_{\mu \in \MK_{S'|S;k}} \Pr(\mu) \frac{\exp\{u_{i}(S')\}}{\sum_{\hat{S} \in \NS_k(\mu,S)}\exp\{u_{i}(\hat{S})\}} \\
& = &\Pfn(S)\sum_{\mu \in \MK_{S|S';k}} \Pr(\mu) \frac{\exp\{\Pfn(S')\}}{\sum_{\hat{S} \in \NS_k(\mu,S)}\exp\{\Pfn(\hat{S})\}}\\
& = &\Pfn(S')\sum_{\mu \in \MK_{S|S';k}} \Pr(\mu) \frac{\exp\{\Pfn(S)\}}{\sum_{\hat{S} \in \NS_k(\mu,S ')}\exp\{\Pfn(\hat{S})\}}\\
& = & \Pfn(S) \Pr(S,S';k)
\end{eqnarray}
Where the particular expression for $Pr(S'|S,\mu)=\frac{\exp\{u_{i}(S')\}}{\sum_{\hat{S} \in \NS_k(\mu,S)}\exp\{u_{i}(\hat{S})\}}$ follows from assumption \ref{asm:gumbel} on the distribution of the error term.
\qed
\end{proof}

\begin{proof}[Theorem \ref{thm:convergencespeed} (p. \pageref{thm:convergencespeed})]
\label{proof:thm-convergencespeed}
Because there is no natural ordering of $\SSn$, use functions as opposed to vectors in the eigenproblem. For $I\subset \{(i,j):i\geq j\}$, define $\e_I : \SSn \rightarrow \R$ as
\begin{equation}
\e_I(S) = \prod_{i\neq j\in I} (-1)^{g_{ij}} \prod_{i=j\in I} (-1)^{a_{ij}}
\end{equation}
with $\e_\emptyset(S) =1$ for all $S$. Next, define 
\begin{equation}
\lambda_{k,I} = \frac{\sum_{i\in\{i:(i,i)\notin I\}}{n-1-|I_i| \choose k-1}}{n{n-1 \choose k-1}}
\end{equation}
where $I_i = \{j:(i,j)\in I, i\neq j\}$

\begin{lemma}
\label{lemma:eigen}
There are $2^{n(n+1)/2}$ pairs of $(\lambda_{k,I},\e_{k,I})$ such that
\begin{enumerate} [topsep=0ex,parsep=0ex,itemsep=0ex,leftmargin=0.7cm,label=(\roman*)]
\item $\sum_S \e_{k,I}(S) \e_{k,I'}(S)=0$ if $I\neq I'$ and $\sum_S \e_{k,I}(S) \e_{k,I}(S)=2^{n(n+1)/2}$
\item For any $S\in \SSn$
\begin{eqnarray}
\sum_{S'} \Pr(S'|S;k)\e_I(S')= \lambda_{k,I} \e_I(S).
\end{eqnarray}
\end{enumerate}
\end{lemma}

\noindent
 
The first part of the lemma is trivial to verify. For the second part, write:
\begin{eqnarray}
\sum \Pr(S'|S)\e_I(S') & = & \sum_{S'} \sum_\mu \Pr(\mu) \Pr(S'|S,\mu) \e_I(S') \\
& = & \sum_{S'} \sum_{\mu\in\{\mu\cap I=\emptyset\}} \Pr(\mu) \Pr(S'|S,\mu) \e_I(S') +  \\ \label{eq:sumvanish}
& & + \sum_{S'} \sum_{\mu\in\{\mu\cap I \neq \emptyset\}} \Pr(\mu) \Pr(S'|S,\mu) \e_I(S') \\ \label{eq:transition}
& =&  \sum_{S'} \sum_{\mu\in\{\mu\cap I=\emptyset\}} \Pr(\mu) \Pr(S'|S,\mu) \e_I(S')
\end{eqnarray}
Terms \eqref{eq:sumvanish} vanish because whenever $\mu\cap I \neq \emptyset$ then $\sum_{S'\in \NS_k(S,\mu)} \Pr(S'|S,\mu) \e_I(S') = 0$, as this summation involves $2^k$ terms and for half of these terms $\e_I(S')=\e_I(S)$ while for the other half $\e_I(S')=-\e_I(S)$, implying that $\sum_{\mu\in\{\mu\cap I \neq \emptyset\}} \sum_{S'\in \NS_k(S,\mu)}\Pr(S'|S,\mu) \e_I(S')$ equals to $0$.

Finally, note that if $\mu\in\{\mu\cap I=\emptyset\}$, i.e. $\mu=\{(i,i),(i,i_1),\ldots (i,i_{k-1})\}\cap I =\emptyset$ then for any $S'\in \NS_k(S,\mu)$ we have that $\e_I(S)=\e_I(S')$ so that for (\ref{eq:transition}) we can write
\begin{eqnarray}
\sum \Pr(S'|S)\e_I(S')
& = & \e_I(S) \Pr(\mu) \sum_{S'} \sum_{\mu\in\{\mu\cap I=\emptyset\}} \Pr(S'|S,\mu) \\
& = & \e_I(S) \frac{1}{n{n-1 \choose k-1}} 
\sum_{\mu\in\{\mu\cap I=\emptyset\}} \sum_{S'} \Pr(S'|S,\mu) \\
& = & \e_I(S) \frac{1}{n{n-1 \choose k-1}} \sum_{i\in\{i:(i,i)\notin I\}} {n-1-|I_i| \choose k-1}
\end{eqnarray}
because, by assumption, $\Pr(\mu) = \frac{1}{n{n-1 \choose k-1}}$ and $\sum_{S'} \Pr(S'|S,\mu)=1$. This completes the proof of lemma \ref{lemma:eigen}. To complete the proof of the theorem note that $\lambda_{k,I}$ are decreasing in $|I|$, so that the (second) largest $\lambda_{k,I}$ is achieved when $I=\{(i,j)\}$ with $i\neq j$.
\qed
\end{proof}

\begin{proof}[Theorem  \ref{thm:equlibriumranking} (p. \pageref{thm:equlibriumranking}] \label{proof:thm:equilibriumranking}
The proof follows immediately from the expression for the stationary distribution obtained in theorem \ref{thm:stationary} and proposition \ref{neksn-representation}.
\qed
\end{proof}

\begin{proof}
[Proposition  \ref{prop:varying_mh} (p. \pageref{prop:varying_mh}]
\label{proof:prop:varying_mh}

For fixed $S,S'\in \SSn$ let $\KK_{S'|S}\subset\{2,3,\ldots, n\}$ be the set of all possible meeting sizes consistent with transition from $S$ to $S'$ of the $k$-PD. Recall that, for fixed $k$, $\MK_{S'|S;k}$ is the set of all possible meetings that may induce transitioning from $S$ to $S'$. The argument bellow follows from lemma \ref{neibourhoods}, together with the observation that $\KK_{S'|S}=\KK_{S|S'}$. Indeed, the unconditional proposal $Q$ from the algorithm in table \ref{table:algorithm} can be written as:

\begin{eqnarray}
Q(S'|S) & = &\sum_{k\in\KK_{S'|S}} p_k(k)\sum_{\mu \in \MK_{S'|S}} \Pr(\mu) \frac{1}{|\NS_k(\mu,S)|} \\
& = &\sum_{k\in\KK_{S|S'}} p_k(k) \sum_{\mu \in \MK_{S|S'}} \Pr(\mu) \frac{1}{|\NS_k(\mu,S)|}\\
& = &\sum_{k\in\KK_{S|S'}} p_k(k) \sum_{\mu \in \MK_{S|S'}} \Pr(\mu) \frac{1}{|\NS_k(\mu,S ')}\\ \notag
& = & Q(S|S')
\end{eqnarray}
\qed
\end{proof}

\section{Implementation details}

This appendix contains details about the data, the sample construction, the parametrization of the model and the estimation. The website \url{www.antonbadev.net/neks} contains additional details including the implementation code and an online appendix with robustness analysis.

\subsection{Add Health Data}
This research uses data from Add Health, a program project directed by Kathleen Mullan Harris and designed by J. Richard Udry, Peter S. Bearman, and Kathleen Mullan Harris at the University of North Carolina at Chapel Hill, and funded by grant P01-HD31921 from the Eunice Kennedy Shriver National Institute of Child Health and Human Development, with cooperative funding from 23 other federal agencies and foundations. Special acknowledgment is due Ronald R. Rindfuss and Barbara Entwisle for assistance in the original design. Information on how to obtain the Add Health data files is available on the Add Health website (http://www.cpc.unc.edu/addhealth). No direct support was received from grant P01-HD31921 for this analysis.

\subsection{Sample selection and sample statistics}

This research uses data from Wave I of Add Health. The in-home questionnaire contains $44$ sections collecting a wide array of information about adolescents. In particular, the data contain information about adolescents' friendship networks. Each respondent is asked to nominate up to five of her best male and female friends. If individual A nominates individual B as a friend, this does not imply that B nominates A. 
 Because in the proposed model a friendship nomination involves \emph{consent}, a friendship presumes that both individuals have nominated each other as friends.%
\footnote{In addition to the in-home interview from Wave I, data on friendship are available from the in-school and Wave III interviews. However, the in-school questionnaire itself does not provide information on important dimensions of an individual's socio-economic and home environment, such as student allowances, parental education, and parental smoking behaviors. On the other hand, during the collection of the Wave III data, the respondents were not in high school any more. For more details on Add Health research design, see \url{www.cpc.unc.edu/projects/addhealth/design} }

In addition to the friendship network data, I use demographic data for the adolescents (age, gender, grade, and race), for their home environments (presence of smoker in the household, pupil's income and allowances, and mother's education), and data for their smoking behavior. The adolescent's smoking status is deduced from the question, ``During the past $30$ days, on how many days did you smoke cigarettes?'' and if the answer was one or more days, the student's smoking status is set to positive. Because all of the students in the saturated sample were eligible for in-home interview, I have detailed information about student friends as well.

As pointed earlier the schools from the saturated sample (16 schools out of 80) were illegible for exhaustive survey. Since the size of the schools from this sample ranges from $20$ to more than $1500$, the smallest and the largest schools are dropped. Also, a special needs school is dropped for having atypical smoking and friendship patterns. After this still the largest school in the sample enrolls more than 4 times more students compared to the second largest. To maintain sample observations of comparable size (each school is an observation), this school is split into grades $9$, $10$, $11$, and $12$ and, for this school, each grade is treated as a separate network.\footnote{Less than $20\%$ of the friendships are inter-grade so that this split does not affect substantially the friendship network.} Finally, schools with fewer than $100$ students are discarded because such large schools are likely to be very different than the rest.\footnote{Indeed, schools with fewer than $100$ students feature very few friendships (median number of friendships $0.6$) and very low smoking rates (median smoking $0.09$).}  Table \ref{table:descriptive_stats} shows selected descriptive statistics for the estimation sample.


\begin{table}[h]
\label{table:descriptive_stats}
\begin{center}
\caption{Descriptive Statistics for the estimation sample}
\begin{tabular}{lcccc}
\hline \hline
            &    Overall &          Min &       Max &     Median \\ \hline
Students    &       1342 &         110 &        234 &        162 \\ 
Smoking     &       0.41 &        0.12 &       0.54 &        0.44 \\ 
Male        &       0.52 &        0.41 &       0.58 &        0.53 \\ 
Whites      &       0.92 &        0.42 &       0.99 &        0.98 \\ 
Blacks      &       0.05 &        0.00 &       0.45 &        0.00 \\ 
As-Hi-Ot    &       0.03 &        0.00 &       0.13 &        0.02 \\ 
Price       &     164.99 &      137.31 &     220.09 &      160.06 \\ 
Avg income  &      83.90 &       47.25 &     145.85 &       71.55 \\ 
Mom edu     &       0.73 &        0.56 &       0.84 &        0.74 \\ 
HH smokes   &       0.48 &        0.25 &       0.61 &        0.51 \\ 
Num friends  &       0.97 &        0.29 &       1.53 &        0.88 \\ 

\hline
\end{tabular}
\label{table:descriptive_stats}
\end{center}
\fignotetitle{Note:} \fignotetext{The final sample contains students from 8 high schools. Min, max, and median are reported at a school level.}
\end{table}

\subsection{Parametrization and re-parametrizations}
\label{app:parameters}
For the empirical specifications selected parameters in \eqref{payoff} and \eqref{payoff2} are functions of the data. In particular, the utility of smoking is 
\begin{eqnarray}
v(X_i) &= & v_0 + v_{price} p_i \\
& & + v_{hhsmokes}\chi(HHS_i)+ v_{momeduc}\chi(MOMEDUC_i)\\
& & +v_{black}\chi(BLACK_i)+v_{grade9+}\chi(GRADE9P_i)
\end{eqnarray}
and the utility of friendship is
\begin{eqnarray}
w(X_i,X_j) &= & w_0+w_{sex}\chi(sex_i \neq sex_j)\\
& & +w_{grade}\chi(grade_i \neq grade_j) +w_{race}\chi(race_i \neq race_j)  
\end{eqnarray}
Also, there is a term $q_{ijk} g_{ij}g_{jk}g_{ki}$ in which $q_{ijk}=q(X_i,X_j,X_k)=q\chi(grade_i>9)\chi(grade_j>9)\chi(grade_k>9)$. In addition to the above $11$ parameters, \eqref{payoff} and \eqref{payoff2} have the externalities' parameters $\phi$, $\phi_S$, and $\phi_N$.

In table \ref{tab:estimates}, the parameters have been transformed for ease of interpretation as follows. Instead of $v_0$, I report the baseline probability of smoking $\theta_1=\frac{e^{v_0}}{1+e^{v_0}}\in[0,1]$. Next, the baseline number of friends is $\theta_8=(n-1)\frac{e^{w_0}}{1+e^{w_0}}\in[0,n-1]$ where $n$ is the size of the network. Also some parameters have been re-parametrized as marginal probabilities in ppt (in table \ref{tab:estimates} indicated as $MP$) or as relative marginal probabilities in pct (in table \ref{tab:estimates} indicated as $MP\%$). For example:\footnote{Note that the reparametrization is bijective so that it does not affect the estimation.}
\begin{eqnarray}
\frac{e^{v_0+v_{hhsmokes}}}{1+e^{v_0+v_{hhsmokes}}}-\frac{e^{v_0}}{1+e^{v_0}}=\theta_3\\
\frac{e^{w_0+w_{diffsex}}}{1+e^{w_0+w_{diffsex}}}:\frac{e^{w_0}}{1+e^{w_0}}=1+\theta_9
\end{eqnarray}

\subsection{Priors and Markov chain parameters}

\begin{table}[t]
\caption{Parameters of the prior distributions}
\label{table:priors}
\begin{center}
\begin{tabular}{llccccc}
\hline \hline
\multicolumn{4}{c}{\textit{Utility of smoking}}  \\
    &           &   Prior &   Prior  & Posterior     & $90\%$       \\
    & Parameter &   mean  &   StD    & mean (median) & Credible set \\ \hline
  1 &Baseline probability of smoking          & 0.20    &   0.10 & 0.18   (0.14) & [0.15, 0.22] \\ 
  2 &Price                         $\times 100$ & -0.50    &   1.00 & -0.24   (-0.61) & [-0.48, -0.01] \\ 
  3 &Mom edu (HS\&CO)$^{MP}$                  & -0.05    &   0.05 & -0.05   (-0.07) & [-0.07, -0.03] \\ 
  4 &HH smokes                                & 0.10    &   0.10 & 0.14   (0.09) & [0.11, 0.17] \\ 
  5 &Grade 9+$^{MP}$                          & 0.20    &   0.20 & 0.16   (0.08) & [0.11, 0.20] \\ 
  6 &Blacks$^{MP}$                            & -0.20    &   0.20 & -0.31   (-0.38) & [-0.37, -0.26] \\ 
  7 &$30\%$ of the school smokes$^{MP}$       & 0.05    &   0.10 & 0.05   (0.01) & [0.03, 0.08] \\ 

\\
\multicolumn{4}{c}{\textit{Utility of friendships}} \\
&           &   Prior &   Prior  & Posterior     & $90\%$       \\
& Parameter &   mean  &   StD    & mean (median) & Credible set \\ \hline
  8 &Baseline number of friends               & 3.00   &   2.00 & 3.40   (2.70) & [2.88, 3.88] \\ 
  9 &Different sex$^{MP\%}$                   & -0.70   &   0.50 & -0.72   (-0.80) & [-0.77, -0.66] \\ 
 10 &Different grades$^{MP\%}$                & -0.70   &   0.50 & -0.89   (-0.93) & [-0.92, -0.86] \\ 
 11 &Different race$^{MP\%}$                  & -0.50   &   0.50 & -0.39   (-0.61) & [-0.56, -0.24] \\ 
 12 &Cost/Economy of scale                    & 0.00   &   0.50 & -0.22   (-0.25) & [-0.24, -0.19] \\ 
 13 &Triangles$^{MP\%}$                       & 0.00   &   2.00 & 1.22   (0.91) & [0.98, 1.45] \\ 
 14 &$\phi_{smoke}^{MP}$                      & 0.05   &   0.05 & 0.05   (0.03) & [0.04, 0.06] \\ 
 15 &$\phi_{nosmoke}^{MP}$                    & 0.05   &   0.05 & 0.04   (0.03) & [0.03, 0.05] \\ 

\hline
\end{tabular}
\label{table:descriptive_stats}
\end{center}
\fignotetitle{Note:} \fignotetext{All prior distributions are normals.}
\end{table}

All priors are set to normal distributions with parameters displayed in table \ref{table:priors}. The other parameters of the algorithm from table \ref{table:algorithm} are as following. The size of the posterior sample is $T=10^5$ from which the first $20\%$ are discarded. The size of the interior loop, from steps $4-12$, is $R=10^3$ for each network. The proposal for $\theta'$ in step 2 is a random walk. The process $k$ is a mixture of two processes: with $75\%$ $k$ is small, i.e. $k=2$ and with $25\%$ it is drawn from discrete uniform on $\{2, \ldots, n-1\}$. Once $k$ is fixed, the state $S'$ in step 8 is drawn from uniform in the permissible neighborhood. In addition, with small probability ($0.05$) a large step is proposed where $S'=1-S$ and $A'=1-A$.

\section{Background on tobacco smoking}

Tobacco is the single greatest preventable cause of death in the world today.\footnote{The World Health Organization, \emph{Report on the Global Tobacco Epidemic} ($2008$). The statistics for the U.S. are compiled from reports by the Surgeon General ($2010$), National Center for Health Statistics ($2011$), and Monitoring the Future (2011).} In the United States alone, cigarette smoking causes approximately $443,000$ deaths each year (accounting for one in every five deaths) and imposes an economic burden of more than $\$193$ billion a year in health care costs and loss of productivity. Approximately $1$ million young people under $18$ years of age start smoking each year; about $80\%$ of adults who are smokers started smoking before they were $18$ \citep{kessler_etal_96,liang_chaloupka_nichter_clayton_01}. Despite an overall decline in smoking prevalence from $2005$ to $2010$, when the percentage of current smokers decreased from $20.9\%$ to $19.3\%$, the reduction in teen smoking has been less pronounced. In fact, the proportions of $8$th and $10$th graders who smoke increased slightly in $2010$. As with many human behaviors, social interactions (peer influence) have often been pointed to as a major driving force behind adolescent smoking choices.

\section{Additional plots and tests}


    \begin{table}[!h]
    \caption{Pairwise tests of the posteriors for the price parameter under different estimation scenarios}
    \label{table:ctrf-posteriorPrice-tests}
    \begin{center}
    \begin{tabular}{lccccccc}
    Estimation & \multirow{2}{*}{Model} & \multirow{2}{*}{Fixed net}&\multirow{2}{*}{No net data}&\multirow{2}{*}{No PE}&\multirow{2}{*}{No tri}&\multirow{2}{*}{No cost} \\
    scenarios &  \\ \hline \hline
    Model        & 1.00 (1.00) \\ 
Exog net     & 0.00 (0.00) & 1.00 (1.00) \\ 
No net data  & 0.00 (0.00) & 0.00 (0.00) & 1.00 (1.00) \\ 
No PE        & 0.00 (0.00) & 0.00 (0.00) & 0.00 (0.00) & 1.00 (1.00) \\ 

    \hline
    \end{tabular}
    \end{center}
    \fignotetitle{Note:} \fignotetext{
    Each cell compares the posterior distribution of the parameter price between a pair of estimation scenarios.
    The two p-values are from testing a hypothesis of equal means and from testing a hypothesis of equal distributions
    (two-sample Kolmogorov-Smirnov test).}
    \end{table}
    

    \begin{table}[!h]
    \caption{Pairwise tests of the policy effects for different levels of price change}
    \label{table:ctrf-posteriorPrice-tests}
    \begin{center}
    \begin{tabular}{lccccccc}
    Policy & \multirow{2}{*}{20} & \multirow{2}{*}{40}&\multirow{2}{*}{60}&\multirow{2}{*}{80}&\multirow{2}{*}{100}&\multirow{2}{*}{120} \\
    level (dP)&  \\ \hline \hline
    20           & 1.00 (1.00) \\ 
40           & 0.00 (0.00) & 1.00 (1.00) \\ 
60           & 0.00 (0.00) & 0.00 (0.00) & 1.00 (1.00) \\ 
80           & 0.00 (0.00) & 0.00 (0.00) & 0.00 (0.00) & 1.00 (1.00) \\ 
100          & 0.00 (0.00) & 0.00 (0.00) & 0.00 (0.00) & 0.00 (0.00) & 1.00 (1.00) \\ 
120          & 0.00 (0.00) & 0.00 (0.00) & 0.00 (0.00) & 0.00 (0.00) & 0.00 (0.00) & 1.00 (1.00) \\ 

    \hline
    \end{tabular}
    \end{center}
    \fignotetitle{Note:} \fignotetext{
    Each cell compares the policy effects for a pair of price changes.
    The two p-values are from testing a hypothesis of equal means and from testing a hypothesis of equal distributions
    (two-sample Kolmogorov-Smirnov test).}
    \end{table}
    

    \begin{table}[!h]
    \caption{Pairwise tests of the response of the overall smoking to same-race caps}
    \label{table:ctrf-schoolcomposition-tests}
    \begin{center}
    \begin{tabular}{cccccccc}
    Same-race & \multirow{2}{*}{0} & \multirow{2}{*}{10}&\multirow{2}{*}{20}&\multirow{2}{*}{30}&\multirow{2}{*}{40}&\multirow{2}{*}{50} \\
    cap ($\%$) &  \\ \hline \hline
      0 & 1.00 (1.00) \\ 
 10 & 0.00 (0.00) & 1.00 (1.00) \\ 
 20 & 0.00 (0.00) & 0.00 (0.00) & 1.00 (1.00) \\ 
 30 & 0.00 (0.00) & 0.00 (0.00) & 0.62 (0.98) & 1.00 (1.00) \\ 
 40 & 0.00 (0.00) & 0.00 (0.00) & 0.00 (0.00) & 0.00 (0.00) & 1.00 (1.00) \\ 
 50 & 0.00 (0.00) & 0.00 (0.00) & 0.00 (0.00) & 0.00 (0.00) & 0.69 (0.69) & 1.00 (1.00) \\ 

    \hline
    \end{tabular}
    \end{center}
    \fignotetitle{Note:} \fignotetext{Each cell examines the change in overall prevelance between a pair of scenarios (same-race caps).
    The two p-values are from testing a hypothesis of equal means and from testing a hypothesis of equal distributions
    (two-sample Kolmogorov-Smirnov test). 
    For example, both tests cannot reject the null (of equal means and equal distributions) 
    of the overall smoking between a same-race cap of 40\% and a same-race cap of 50\% (p-value $0.37 (0.95)$).
    For all other cases the policy induces statistically significant changes in the overall smoking.}
    \end{table}

\begin{figure}[!h]
	\caption{Overall smoking (schools Black and White) under different same-race caps}
	\label{fig:schcomposition}
	\centering
	\includegraphics[scale=0.8]{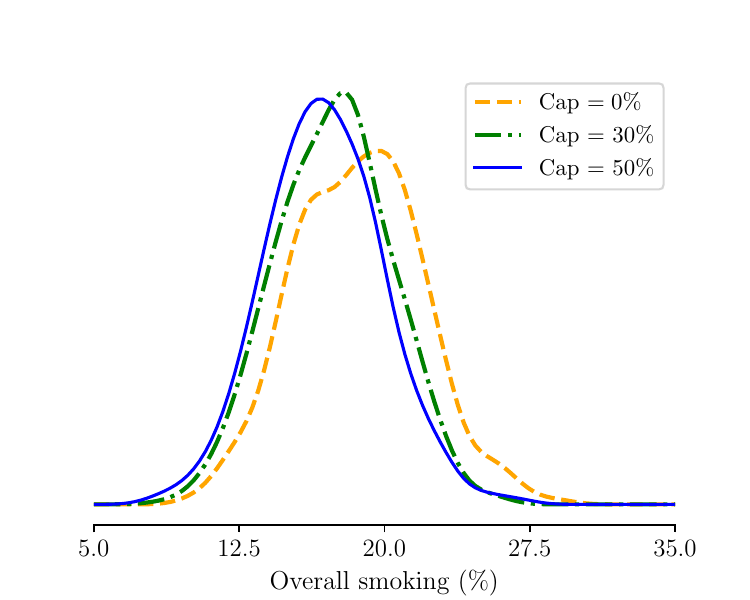}
\end{figure}


\clearpage
\newpage
\bibliographystyle{aer}
\addcontentsline{toc}{chapter}{Bibliography}
\bibliography{bib01}

\end{document}